\documentclass[11pt,oneside,letterpaper]{article}
\pdfoutput=1
\usepackage{amssymb}
\usepackage{graphicx}
\usepackage{lipsum}
\usepackage{slashed}
\usepackage{fancyhdr}
\usepackage{amsmath}
\usepackage{xcolor}
\usepackage{graphicx}
\usepackage[colorlinks=true, linkcolor  = blue,urlcolor=blue,citecolor=violet]{hyperref}
\usepackage{subfigure}
\usepackage[export]{adjustbox}
\usepackage{wrapfig}
\usepackage[margin=2.8cm]{geometry}
\usepackage{setspace}
\usepackage{multirow,makecell}
\usepackage{tabularx}
\usepackage{nicefrac}
\usepackage[font=footnotesize,labelfont=bf]{caption}

\def\({\left(} \def\){\right)}
\def\[{\left[} \def\]{\right]}

\newcommand{\ie}{{\it i.e.,}\ }

\newcommand{\be}{\begin{equation}}
\newcommand{\ee}{\end{equation}}
\newcommand{\bea}{\begin{eqnarray}}
\newcommand{\eea}{\end{eqnarray}}

\def\a{\alpha}

\catcode`\@=12

\renewcommand{\eqref}[1]{(\ref{#1})}

\numberwithin{equation}{section}

\begin{document}

\thispagestyle{empty}

\vspace*{2.5cm}
\begin{center}
{\LARGE\textsc{De Sitter Horizons $\&$ Holographic Liquids}}\\

\vspace*{1.7cm}
\normalsize{Dionysios Anninos$^1$, Dami{\'a}n A. Galante$^2$, and Diego M. Hofman$^2$}
\end{center}
$\,$\newline
{\footnotesize{$^1$ Department of Mathematics, King's College London, Strand, London WC2R 2LS, UK \newline
$^2$ Institute for Theoretical Physics Amsterdam $\&$ $\Delta$ Institute for Theoretical Physics,
University of Amsterdam, Science Park 904, 1090 GL Amsterdam, The Netherlands}}
 
\vspace*{0.6cm}

\vspace*{1.5cm}
\begin{abstract}
\noindent

We explore asymptotically AdS$_2$ solutions of a particular two-dimensional dilaton-gravity theory. In the deep interior, these solutions flow to the cosmological horizon of dS$_2$. We calculate various matter perturbations at the linearised and non-linear level. We consider both Euclidean and Lorentzian perturbations. The results can be used to characterise the features of a putative dual quantum mechanics. The chaotic nature of the de Sitter horizon is assessed through the soft mode action at the AdS$_2$ boundary, as well as the behaviour of shockwave type solutions. 

\end{abstract}

\newpage
\setcounter{page}{1}
\pagenumbering{arabic}

\tableofcontents

\onehalfspacing

\section{Introduction}

The AdS/CFT correspondence has provided important evidence in support of the idea that the horizon of a black hole, at the microscopic level, comprises a large number of strongly interacting degrees of freedom in a liquid-like dissipative state. The role of AdS is to provide the horizon with a non-gravitating boundary from which the characteristic features of the holographic liquid can be probed. The simplest way to do so is through the evaluation of correlations of low energy local operators. The goal of this paper is to make progress toward the hypothesis that the cosmological horizon of an asymptotically de Sitter universe is itself, microscopically, a holographic liquid. Several thermodynamic features of the de Sitter horizon have been known since the classic work of Gibbons and Hawking \cite{Gibbons:1977mu}. What is missing so far, is a framework to bridge the gap between the thermodynamic and microscopic, in the same spirit that AdS/CFT bridges the gap between the old literature on black hole thermodynamics and the modern perspective of a holographic liquid. 

What makes the de Sitter problem challenging is the absence of a spatial AdS boundary, or more generally some non-gravitating region of spacetime, from which to probe the de Sitter horizon. To address this, we construct a phenomenological gravitational theory which contains asymptotically AdS solutions with a region of de Sitter in the deep interior. Our approach is inspired, to an extent, by analogous approaches applying the framework of AdS/CFT to problems in condensed matter \cite{Hartnoll:2009sz}. 
Though this approach is incomplete, we believe that bringing the question of the de Sitter horizon to the standards of AdS/CMT is a step in the forward direction. Ultimately, a successful approach will require a microscopic completion. 

Concretely, we construct a class of two-dimensional gravitational theories admitting solutions which interpolate between an AdS$_2$ boundary with a dS$_2$ horizon in the deep interior \cite{Anninos:2017hhn}. Having done so, we probe the de Sitter horizon using the available tools of AdS/CFT. There are several reasons to work in two-dimensions. Though simpler, the dS$_2$ horizon shares many features with its higher dimensional cousin, including the characteristics of its quasinormal modes \cite{LopezOrtega:2006my} and the finiteness of the geometry. Also, dS$_2 \times S^2$ appears as solution of Einstein gravity with $\Lambda>0$ \cite{nariai} and hence dS$_2$ is directly relevant to four-dimensional de Sitter. Furthermore, recent progress in our microscopic understanding of AdS$_2$ holography \cite{Sachdev:1992fk,parcollet,kitaev,Maldacena:2016hyu,Sachdev:2015efa,Polchinski:2016xgd,Anninos:2013nra,Anninos:2016szt} may guide us in constructing a microscopic model dual to the interpolating geometry. 

Embedding an inflating universe in an AdS$_{d+1}$ spacetime with $d>1$ was previously considered in the interesting works of \cite{Freivogel:2005qh,Lowe:2010np}. There, essentially due to the Raychadhuri equation combined with the null energy condition, it was found  the de Sitter region lived within the Schwarzschild-AdS black hole horizon. It would seem, then, that discovering the de Sitter horizon is at least as complicated as solving the notorious puzzle of the region interior to the horizon. The two-dimensional geometries we consider have the advantage that the de Sitter horizon and its neighbouring static region are causally connected to the AdS$_2$ boundary. The reason we can do this is crucially related to the dimensionality of the AdS$_2$ boundary, and is precisely the one case not considered in previous literature. It is closer in spirit to a `holographic worldline' perspective of the static de Sitter region \cite{Anninos:2011af,vanLeuven:2018pwv}. From this perspective, it is assumed that the dual description of the static de Sitter region is captured by a large $N$ quantum mechanical model, rather than some local quantum field theory in $d>0$ spatial dimensions.

The paper is organised as follows. In section \ref{generalsec} we discuss the dilaton-gravity theories of interest and their solution space, as well as the boundary value problems of interest. The class of solutions admit a $U(1)$ isometry as well as an AdS$_2$ boundary. Of these, one branch is described by a metric which interpolates between an AdS$_2$ boundary and the static patch of dS$_2$. In section \ref{pertsec} we consider matter perturbations at the linearised level. In section \ref{schwartzian} we consider the dynamics of the soft mode residing at the boundary of the interpolating solutions. In section \ref{sec_otoc} we calculate out-of-time-ordered four-point functions stemming from the exchange of the soft mode. We note the absence of exponential Lyapunov behaviour. In section \ref{shocksec} the backreaction of a shockwave pulse on the interpolating geometry is explored. It is noted that in certain circumstances, the horizon can retreat rather than advance toward the AdS$_2$ boundary. In section \ref{discussion}, we conclude with a discussion of the results, their potential holographic implications, and some speculative remarks. Finally, further details for certain calculations can be found in the appendices.

\section{General framework}\label{generalsec}

The theory we consider is described by the following dilaton-gravity Euclidean action:
\begin{equation}\label{theory}
S_E = - \frac{1}{2 \kappa} \int d^2x \sqrt{g} \left( \phi R +V(\phi) \right)  - \frac{1}{\kappa} \int_{\partial\mathcal{M}} du \sqrt{h} \, \phi \, K \,+  S_{m}~,
\end{equation}
where $S_m$ is the action for some matter theory. One can also add to this action a topological term:
\begin{equation}\label{stop}
S_{top} = - \frac{\phi_0}{2}  \int d^2x \sqrt{g}   R   -  {\phi_0} \, \int_{\partial\mathcal{M}} du \sqrt{h} \,  K~,
\end{equation}
where $\phi_0$ is a positive constant which we consider to be large. The Newton constant is given by $\kappa = 8\pi G$. The scalar $\phi$ in (\ref{theory}), or equivalently $\kappa$, may be positive or negative -- what is important is that $\phi_{tot} = ( \phi_0 + \phi/\kappa)$ remain everywhere positive. Hence, throughout the discussion we will assume that $\phi_0 \gg |\phi/\kappa| \gg 1$ and consider both positive and negative values of $\kappa$. The Euclidean geometry lives on a disk topology $\mathcal{M}$ with a circular boundary $\partial\mathcal{M}$. 

The equations of motion for the two-dimensional metric and dilaton read:
\begin{eqnarray}
\nabla_a \nabla_b \phi - g_{ab} \nabla^2 \phi + \frac{g_{ab}}{2}  V(\phi)  &=& - \kappa \,   T_{ab}^{m}~, \label{Tphi} \\
R & = & - V'(\phi) \label{Ricci}~,
\end{eqnarray}
where $T_{ab}^m$ is the stress tensor for the matter fields. There are also the matter equations of motion. We assume further, that the matter theory interacts only with the two-dimensional metric at the classical level. Taking the divergence of equation (\ref{Tphi}) leads to:
\begin{equation}
[\nabla^2 , \nabla_a ] \phi + \frac{1}{2} \, \nabla_a V(\phi)  = 0~. 
\end{equation}
Some algebra reveals that $[\nabla^2,\nabla_a]\phi = R \, \partial_a \phi/2$. Using this, the divergence equation reduces to:
\begin{equation}
R \, \partial_a \phi = - \partial_a V(\phi)~.
\end{equation}
%
In other words, one finds that equation (\ref{Ricci})  is redundant whenever $\partial_a \phi \neq 0$. 
Another useful equation is obtained by taking the trace of (\ref{Tphi}) that leads to:
\begin{equation}\label{nablaphi}
 - \nabla^2 \phi + V(\phi)= -\kappa  \, T^{m}_{ab} \, g^{ab}~.
\end{equation}
Finally, in the absence of matter it is straightforward to check that the equations of motion imply $\xi^a = \epsilon^{a b} \partial_b \phi$ is a Killing vector \cite{Banks:1990mk}.

\subsection{Gauge fixing}

We will be considering tree level features of the theory (\ref{theory}). For this purpose, a useful gauge is the conformal gauge:
\begin{eqnarray}\label{gauge}
ds^2 =  e^{2\omega (\rho,\tau)} (d\rho^2 + d\tau^2) \,,
\end{eqnarray}
where $\tau$ is periodic with period $2\pi$ and the origin of the disk lives at $\rho \to -\infty$. 

Using (\ref{nablaphi}), the dilaton equations become:
\begin{eqnarray}
\partial_{\rho}\partial_\tau\phi -  \partial_\tau\omega\,\partial_\rho\phi - \partial_\rho\omega \, \partial_\tau\phi &=& - \kappa T^m_{\rho\tau}~, \\
\frac{1}{2}\left(\partial_{\rho}\partial_\rho -\partial_\tau \partial_\tau \right) \phi -  \partial_\rho\omega \, \partial_\rho\phi + \partial_\tau\omega \, \partial_\tau\phi  &=& - \frac{\kappa}{2} \left(T^m_{\rho\rho} - T^m_{\tau\tau} \right)~.
\end{eqnarray}
Another gauge which is often convenient is the Schwarzschild gauge:
\begin{equation}\label{schwcoord}
ds^2 = N(r,T)dT^2 + \frac{dr^2}{N(r,T)}~.
\end{equation}
In the Schwarzschild gauge, the Euclidean solution is \cite{Cavaglia:1998xj}:
\begin{equation}\label{schsol}
N(r) = \frac{1}{|\phi_h|} \int_{\text{sign} \, \phi_h}^r dz \, V(\phi(z))~, \quad\quad \phi(r) = |\phi_h| r~.
\end{equation}
The periodic condition $T \sim T+2\pi$ is fixed by requiring a regular Euclidean geometry. The point $r= \text{sign} \, \phi_h$ is the location of the Euclidean horizon. 

Given the Killing vector $\xi^a = \epsilon^{a b}\partial_b \phi$, we will (locally) choose $\omega$ to be solely a function of the $\rho$-coordinate. This dramatically simplifies the matter-less equations, which can be now be solved explicitly. 

\subsection{Choice of potential and background solution}

We will be interested in the class of potentials introduced in \cite{Anninos:2017hhn}. We take $V(\phi)$ to be a non-negative function. Outside some transition region $|\phi| \gtrsim \epsilon$ with $\epsilon$ a small positive number, the potential behaves as $V(\phi) \approx 2|\phi|$. This transition region is not very important, but we assume that $V(\phi)$ is continuous and vanishing at $\phi=0$. Two simple examples are $V(\phi) = 2 |\phi|$ and $V(\phi) = 2 \phi \tanh ( \phi/\epsilon )$. For positive/negative $\phi$ the metric has constant negative/positive curvature. Depending on the value of the dilaton at the origin of the disk, which we denote by $\phi_h$, the metric may have both negative and positive curvature, or purely negative curvature.

For most of the paper we will be interested in the sharp gluing limit with $\epsilon \rightarrow 0$. In the conformal gauge with $\phi_h<0$, the solution for the metric is given by:

\begin{equation}\label{bmetric}
ds^2 = \begin{cases}
\cos^{-2}\rho \left( d\rho^2 + d\tau^2 \right)~, \quad\quad \rho \in (0,\pi/2)~, \\
 \cosh^{-2}\rho\left( d\rho^2 + d\tau^2 \right)~, \quad\quad \rho \in  (-\infty, 0)~.
\end{cases}
\end{equation}
Smoothness requires $\tau \sim \tau + 2\pi$. The dilaton equation is solved by:
\begin{equation}\label{bdilaton}
\phi  = \begin{cases}
  -\phi_h \tan \rho ~, \quad\quad \rho \in(0,\pi/2) \,, \\
 -\phi_h \tanh \rho ~, \quad\quad \rho \in (-\infty,0) \,.
 \end{cases}
\end{equation} 
We will refer to expressions (\ref{bmetric}) and (\ref{bdilaton}) as the interpolating solution. The Euclidean AdS$_2$ boundary lives at $\rho = \pi/2$, whereas for negative values of $\rho$ the metric is the standard metric on a half-sphere. The Euclidean AdS$_2$ piece of the geometry (\ref{bmetric}), between $\rho \in (0,\pi/2)$, is related by analytic continuation to the global Lorentzian AdS$_2$ geometry. It can be viewed as a piece of the hyperbolic cylinder. 

For $\phi_h>0$, the solution has negative curvature everywhere and is given by:
\begin{eqnarray}
ds^2 &=& \sinh^{-2}\rho\left( d\rho^2 + d\tau^2 \right)~, \quad\quad  \rho \in(-\infty,0)~, \label{cmetric}  \\
\phi &=& -\phi_h  \coth \rho~. \label{cdilaton}
\end{eqnarray}
Again, smoothness requires $\tau \sim \tau + 2\pi$. The geometry (\ref{cmetric}) is the hyperbolic disk. It is the Euclidean continuation of the AdS$_2$ black hole geometry.

As we mentioned earlier, we will allow both signs of $\kappa$ in (\ref{theory}). To give some context to this, consider the spherically symmetric sector of Einstein gravity with a positive cosmological constant. In this case, the dimensionally reduced theory in two-dimensions is itself a dilaton-gravity theory. In a particular limit, known as the Nariai limit, the dilaton potential is approximately linear.  The Euclidean solution is given by the two-sphere with a running dilaton. The dilaton increases from one pole to the other. Viewing the two-sphere as two hemispheres joined at the equator, we see that one hemisphere has an increasing dilaton profile towards the pole, whereas the other has a decreasing dilaton profile. Switching from one sign to the other switches the sign of $\kappa$ in the effective two-dimensional theory. In appendix \ref{wrongsign} we discuss a qualitatively similar phenomenon in the context of pure Jackiw-Teitelboim theory \cite{JTgravity}. In appendix \ref{gammaSch} we discuss a broader family of dilaton potentials where the potential changes behaviour at some value $\phi_0$ and we relax the assumption $V(\phi)>0$. 

\subsubsection*{\it Lorentzian continuation}

\begin{figure}[h!]
        \centering
                \includegraphics[scale=0.3]{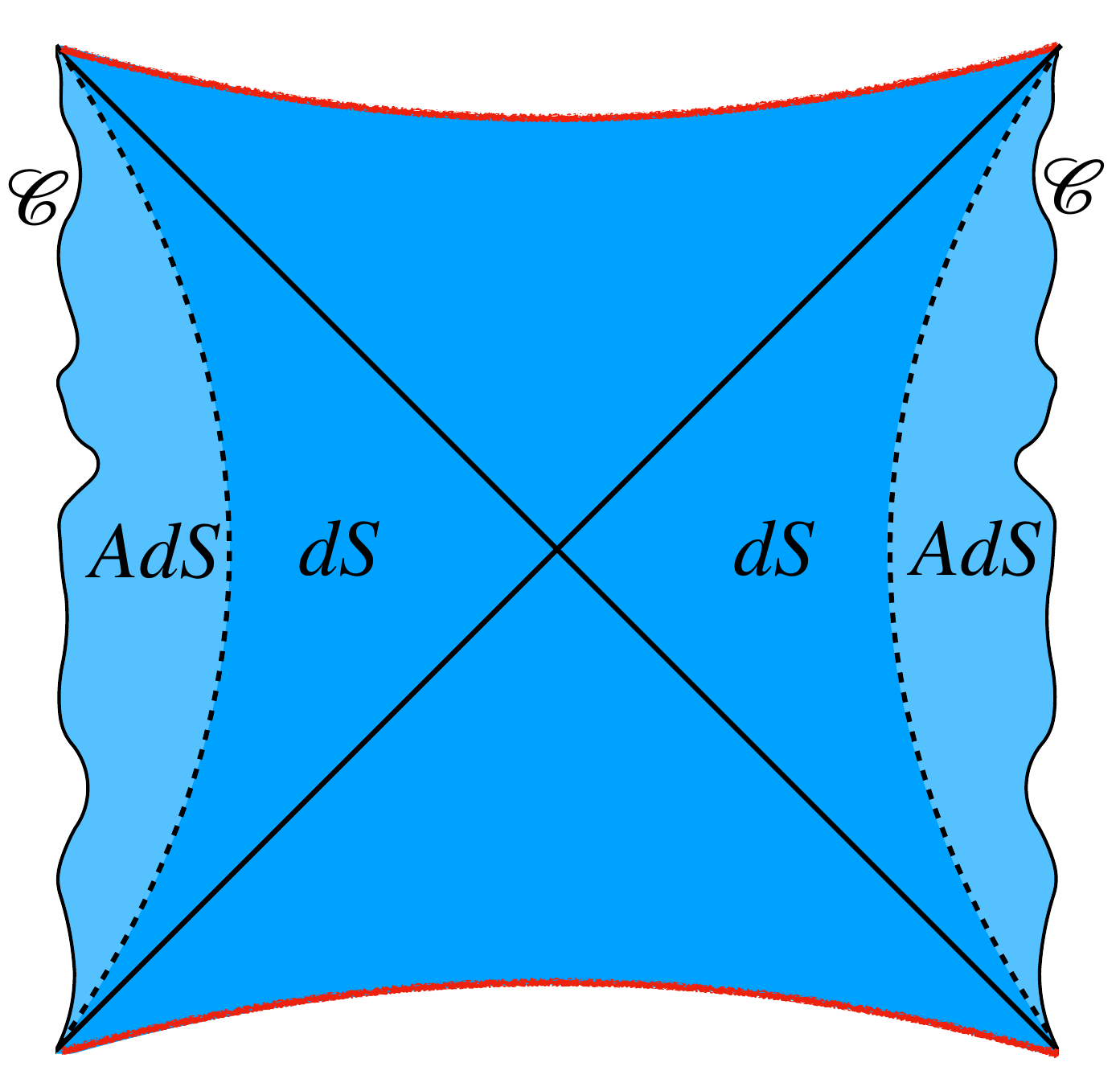}
                 \caption{{\footnotesize Penrose diagram for the interpolating solution. The dashed lines interpolate between a negative and a positive curvature region, that are coloured in light and darker blue, respectively. $\mathcal{C}$ is the boundary curve close to the AdS boundaries. Inside the horizons, the geometry is locally dS$_2$, but depending on whether $\kappa$ is positive or negative, the dilaton behaves as in the interior of a dS black hole or as in the dS cosmological patch.}}
\label{fig_background}
\end{figure}

As a final note, we mention that the Euclidean solutions can be continued to Lorentzian solutions. This continuation can be done in several ways. The simplest continuation takes $\tau \to i t$. The resulting solution is static and has a horizon at the value where the Euclidean geometry smoothly capped off, namely $\rho \to -\infty$. The boundary is that of an asymptotically Lorentzian AdS$_2$ geometry. We can extend the geometry beyond the horizon. The interior of the horizon is either locally dS$_2$ or the interior of the AdS$_2$ black hole, depending on the sign of $\phi_h$. The Penrose diagram for the interpolating geometry is shown in figure \ref{fig_background}.

One could also imagine analytically continuing $\rho$ instead. In that case, one obtains a cosmological type solution and the dilaton becomes time-dependent. If $\tau$-remains periodically identified upon continuing $\rho$, the interpolating cosmological geometry will have a big bang/crunch type singularity and asymptote to the future/past boundary of global dS$_2$. Though interesting in their own right, these solutions have compact Cauchy surfaces and are thus not asymptotically AdS$_2$. We leave their consideration for future work.

\subsection{Boundary value problem}

We now discuss several boundary conditions for the Euclidean dilaton-gravity theory (\ref{theory}). 


The Dirichlet problem is given by specifying the value of the dilaton $\phi_b(u)$ and induced metric $h(u)$ at $\partial \mathcal{M}$. Here, $u$ is a compact coordinate that parameterises points on $\partial\mathcal{M}$. In the Weyl gauge:
\begin{equation}\label{induced}
h(u) = e^{2\omega(\rho(u),\tau(u))} \left( (\partial_u \tau(u) )^2 + ( \partial_u \rho(u) )^2 \right)~, \quad \phi_b(u)  = \phi(\rho(u),\tau(u))~,
\end{equation}
where $\mathcal{C} = \{ \tau(u),\rho(u)\}$ is the curve circumscribing $\partial \mathcal{M}$. The curve $\mathcal{C}$ is not independent data. Rather, it is fixed by the particular solution to the Dirichlet problem.
 Requiring the variation of the action to vanish under these conditions leads to the addition of the usual Gibbons-Hawking boundary term. If the boundary value $\phi_b(u)$ is positive everywhere, then $\mathcal{C}$ circumscribes part of the negatively curved geometry. If, moreover, $h(u)$ and $\phi_b(u)$ scale with some large parameter $\Lambda$, such that $\{\sqrt{h},\phi_b\}= \Lambda \{ \sqrt{\tilde{h}},\tilde{\phi}_b\}$ in the limit $\Lambda \to \infty$, we can approximate the Gibbons-Hawking boundary term by:
\begin{equation}\label{bdyaction}
S_{bdy}[\tau(u)]  \approx - \frac{1}{\kappa}   \int du \sqrt{\tilde{h}} \, \tilde{\phi}_b \left( \Lambda^2 + \text{sign}\, {\phi}_h \, \frac{(\partial_u \tau)^2}{2 \tilde{h}(u)} +  \text{Sch}_{\tilde{h}}\left[\tau(u),u  \right] \right)~.
\end{equation}
The above action and its properties will be developed further in section \ref{schwartzian}. The covariant Schwarzian action $\text{Sch}_{\tilde{h}}\left[\tau(u),u\right]$ is obtained by taking the standard Schwarzian $\text{Sch}\left[\tau(u),u\right]$ and coupling it to a non-trivial metric $\tilde h(u)$. We will write the more general expression later on in (\ref{diegoformula}). For now, we fix $\tilde h(u) = 1$ and state the familiar expression:
\begin{equation}
\text{Sch}\left[\tau(u),u\right] = \frac{\partial_u^3 \tau}{\partial_u \tau} - \frac{3}{2} \left( \frac{\partial_u^2 \tau}{\partial_u \tau} \right)^2~.
\end{equation}
The reason for the $\text{sign}\, \phi_h$ in (\ref{bdyaction}) is that the background solution has a different radial slicing depending on whether or not there is a two-sphere in the interior. In particular, for $\phi_h>0$ the asymptotic AdS$_2$ clock is that of the Euclidean AdS$_2$ black hole, i.e. the isometric direction of the hyperbolic disk. When $\phi_h<0$, the asymptotic AdS$_2$ clock is that of Euclidean global AdS$_2$ with periodically identified time, i.e. the hyperbolic cylinder. Consequently the function $\tau(u)$ must be a map from $S^1$ to $S^1$. 

Using the equations of motion, we can obtain an expression for the on-shell Euclidean action for the Dirichlet problem:
\begin{equation}\label{onshell}
- S_{cl} = \frac{1}{\kappa} \int du \sqrt{h(u)} \, \phi_b(u) K(u) + \frac{1}{2\kappa} \int d^2 x \sqrt{g} \left( V(\phi) - \phi \, \partial_\phi V(\phi) \right)~.
\end{equation}
It is straightforward to evaluate the above action in the Schwarzschild gauge (\ref{schwcoord}). The second term in (\ref{onshell}) is non-vanishing only near the region where $V(\phi)$ is non-linear. 
This region can be made parametrically small, such that the dominant contribution to the on-shell action is from the boundary term.\footnote{More generally, one could also consider adding additional local boundary terms \cite{Grumiller:2007ju}.} For instance, take $V(\phi) \approx \zeta^{-1}  \phi^2$ near the transition region around $\phi=0$, and let the width of the region be $2\epsilon$. Then, it can be shown, using the general solution (\ref{schsol}), that:
\begin{equation}\label{onshell2}
-S_{int} = \frac{1}{2\kappa} \, \int d^2 x \sqrt{g} \left( V(\phi) - \phi \partial_\phi V(\phi) \right) = \frac{\alpha\epsilon^3}{3\kappa\zeta}~, \quad\quad \phi_h \ll -\epsilon~,
\end{equation}
where $\alpha=4\pi/V(\phi_h)$. For small enough $\epsilon$, the dominant contribution to the on-shell action comes from the boundary term (\ref{bdyaction}). As a concrete example, one can consider $V(\phi) = 2 \phi \tanh \phi/\epsilon$, for which an exact calculation can be performed. In this case, $\zeta=\epsilon$, and one can indeed check that $S_{int}\sim \epsilon^2$. Note that this contribution is not present if $\phi_h \gg \epsilon$.

The Neumann problem is given by specifying the value of the conjugate momenta $\pi_{\phi_b}(u)$ and $\pi_{h}(u)$ along a prescribed closed boundary curve $\mathcal{C} = \{\tau(u),\rho(u)\}$ parameterized by some coordinate $u$. Explicitly, the conjugate (Euclidean) momenta to the boundary metric and dilaton are:
\begin{equation}
\kappa \, \pi_{\phi_b} = -K~, \quad\quad \kappa \, \pi_{h} = - \frac{n^a \partial_a \phi}{2 h}  ~.
\end{equation}
Here $n^a$ is a unit vector that is normal to $\mathcal{C}$. Since now the dilaton and induced metric are allowed to vary along $\mathcal{C}$, we may find saddle point solutions for which the metric is nowhere AdS$_2$. As a simple example, if we fix $\{\pi_h,\pi_{\phi_b} \} = \{-\nicefrac{1}{2 \kappa}\,,0\}$ as our boundary condition, the solution is given by the half-sphere. 

In addition to the Dirichlet and Neumann boundary conditions, we may also consider mixed Dirichlet-Neumann boundary conditions or even conditions on linear combinations of the momenta and boundary values of the fields. The choice depends on the physics of interest. Not all boundary data $\{\phi_b(u),h(u)\}$ admit a Euclidean solution which is both smooth and real. For instance, if $\phi_b$ is $u$-independent and negative, and $h$ is $u$-independent and sufficiently large, one cannot realise a smooth and real Euclidean saddle. There may be complex saddles that are allowed however. In recent literature relating two-dimensional gravity to the SYK model (see for example \cite{Maldacena:2016upp}), the Dirichlet problem is considered for the two-dimensional bulk theory, and we will mostly focus on this choice.

As a concrete example, we consider the Dirichlet problem with $u$-independent boundary values $\phi_b \gg 1$ and $h \gg 1$. There are two-solutions satisfying these boundary conditions. For $\phi_h<0$ it is (\ref{bmetric}), (\ref{bdilaton}) with $-\phi_h \tan\rho_c = \phi_b$ and $\cos^{-2}\rho_c=h$. For $\phi_h>0$ it is (\ref{cmetric}), (\ref{cdilaton}) with $-\phi_h \coth\rho_c = \phi_b$ and $\sinh^{-2}\rho_c=h$. The boundary lies at a constant $\rho=\rho_c$ surface near the AdS$_2$ boundary, with $\tau=u$. The configuration with least action depends on the sign of $\kappa$. For $\kappa>0$ ($\kappa<0$) the dominant configuration has $\phi_h>0$ ($\phi_h<0$).

In the remainder, we explore the behaviour of the background solutions upon turning on various matter sources in the Dirichlet problem and we will make short comments on the Neumann problem when appropriate. 

\section{Perturbative analysis}\label{pertsec}

In this section we consider the effect of a small matter perturbation. The goal is to gain some understanding as to how the geometry responds to matter perturbations when the background has an interpolating region. Unlike the case of the Jackiw-Teitelboim model \cite{JTgravity} which has a linear dilaton potential which fixes the internal geometry entirely, our perturbations will also involve the metric. 

We consider turning on some matter content with non-trivial boundary profile. This will interact with the graviton and dilaton fields with strength $\kappa$, such that the perturbative expansion is one in small $\kappa$. The geometry communicates with the matter through the effect of the matter on the dilaton. We choose our background solution to be the interpolating geometry (\ref{bmetric}) with dilaton profile (\ref{bdilaton}), which assumes the sharp-gluing limit where $\epsilon = 0$, i.e. $V(\phi) = 2 |\phi|$. Given the background $U(1)$ isometry in the $\tau$-direction, it is convenient to consider the Fourier modes of the linearised fields. As matter, we consider a complex, massless, free scalar $\chi$ with action:
\begin{equation}\label{matterS}
S_\chi = \int d^2 x \sqrt{g} \, g^{ab} \, \partial_a \bar{\chi} \, \partial_b \chi~.
\end{equation}
In the conformal gauge, the general solution that is well behaved in the interior is:
\begin{equation}
\chi(\rho,\tau) = \sum_{m \in \mathbb{Z}} h_m e^{|m| (\rho-\pi/2) + i m \tau}~, \quad\quad h_m \in \mathbb{C}~.
\end{equation}
At the AdS$_2$ boundary, where $\rho = \pi/2$, the Fourier modes of the boundary profile for the matter field are $h_m$. The $\rho\tau$-component of the matter stress-tensor will be most relevant for our calculations. We first solve the general linearised equations, and then consider particular boundary conditions.

\subsection{Linearised equations}

To leading order in a small matter field expansion, we must solve for the fluctuation $\delta \phi(\rho,\tau)$ of the background dilaton $\bar{\phi}(\rho)$ and the fluctuation $\delta \omega(\rho,\tau)$ of the background conformal factor $\bar{\omega}(\rho)$. The perturbed geometry can be expressed as:
\begin{equation}
ds^2 = e^{2\left(\bar{\omega}(\rho) + \delta \omega(\rho,\tau)\right)} \left( d\rho^2 + d\tau^2 \right)~.
\end{equation}
We obtain the following linearised equations:
\begin{eqnarray}
\partial_\rho \partial_\tau \delta \phi - \partial_\rho \bar{\omega} \partial_\tau \delta \phi -  \partial_\tau \delta\omega \partial_\rho \bar{\phi}   &=& - \kappa \, T_{\tau\rho}^\chi~, \label{deltaphi}\\
 - 2 e^{-2\bar{\omega}} \left(\partial_\rho^2 + \partial_\tau^2 \right) \delta \omega - 2 \bar{R} \,  \delta \omega &=& -   \partial_\phi^2 V(\bar{\phi}) \, \delta \phi  \label{deltaomegaeq}~,
\end{eqnarray}
where now $T_{\tau\rho}^\chi$ is the stress tensor corresponding to the scalar $\chi$. The inhomogeneous solution to (\ref{deltaphi}) can be expressed as an integral:
\begin{equation}\label{integral}
\delta \phi_{inh}(\rho,\tau) =  e^{\bar{\omega}(\rho)}   \int_{-\infty}^\rho  d\rho' e^{-\bar{\omega}(\rho')}  \left( -\kappa \int^\tau d\tau' T_{\rho\tau}^\chi(\rho',\tau') +  \partial_{\rho'} \bar{\phi}(\rho')   \delta \omega(\rho',\tau)   \right)~.
\end{equation}
Substituting (\ref{integral}) into (\ref{deltaomegaeq}) we find:
\begin{equation}
\left( - 2 e^{-2\bar{\omega}(\rho)} \left(\partial_\rho^2 + \partial_\tau^2 \right) - 2 \bar{R} \right) \delta \omega(\tau,\rho)  + \partial_\phi^2 V(\bar{\phi}) \, \delta \phi_{inh}(\rho,\tau)    = 0~.
\end{equation}
For $V(\phi) = 2|\phi|$ we have $\partial_\phi^2 V(\phi)=4\delta(\phi)$. Away from the $\rho=0$ interpolating region, we must solve the free wave-equation:
\begin{equation}\label{deltaR}
\left( -  e^{-2\bar{\omega}(\rho)} \left(\partial_\rho^2 + \partial_\tau^2 \right) -  \bar{R} \right) \delta \omega(\rho,\tau)  = 0~.
\end{equation}
The above equation states that the linearised correction to the Ricci scalar must vanish. Its solution corresponds to a linearised diffeomorphism that preserves the conformal gauge. For either $\rho>0$ or $\rho<0$, (\ref{deltaR}) has two solutions. In the $\rho<0$ region, we must solve:
\begin{eqnarray}
\left( \cosh^2\rho \left(\partial_\rho^2 + \partial_\tau^2 \right) + 2 \right) \delta \omega_-(\rho,\tau) = 0~.
\end{eqnarray}
We keep the solution that is non-singular at $\rho = -\infty$. The Fourier modes are then found to be:
\begin{equation}
\delta \omega_-(\rho,q) = \alpha_q \, e^{|q| \rho} \, \left(-|q|+\tanh \rho \right)~, \quad\quad q \in \mathbb{Z}~,
\end{equation}
with $\a_q = \alpha_{-q}^*$. For $\rho>0$, we must solve:
\begin{eqnarray}
\left( \cos^2\rho \left(\partial_\rho^2 + \partial_\tau^2 \right) - 2 \right) \delta \omega_+(\rho,\tau)    = 0~.
\end{eqnarray}
Interestingly, the above equation is that of a free scalar in AdS$_2$ with $\Delta=2$. The two solutions are given by:
\begin{equation}
\delta \omega_+ (\rho,q) = \beta_q \, e^{|q| \rho} (|q| + \tan \rho) + \gamma_q \, e^{-|q| \rho} (|q| - \tan \rho)~, \quad\quad q \in \mathbb{Z}~, \label{deltaomegaplus}
\end{equation}
with $\beta_q = \beta_{-q}^*$ and $\gamma_q = \gamma_{-q}^*$~.
We are now in a position to compute the integral (\ref{integral}). For the jumping condition, we need the integral at $\rho=0$.
In order to fix some of the coefficients of the solution, we must consider matching $\delta\omega_-(\rho,\tau)$ and $\delta \omega_+(\rho,\tau)$ across $\rho=0$. The first matching condition is continuity of the metric at $\rho=0$, so that
\begin{eqnarray}
\delta\omega_- (0,\tau) = \delta\omega_+(0,\tau) \,.
\end{eqnarray} 
From this, it is straightforward to get
\begin{equation}
\alpha_q = - {\left(\gamma_q + \beta_q\right)} \,. \label{alphas}
\end{equation} 
The second matching condition relates the radial derivatives of $\delta\omega(\rho,\tau)$ across $\rho=0$. From $\partial_\phi^2 V(\phi)=4\delta(\phi)$, one obtains the following jump condition:
\begin{equation}
\left(\partial_\rho   \delta \omega_+(\rho,\tau) -\partial_\rho \delta \omega_-(\rho,\tau)  \right) \Big |_{\rho=0} =  - \frac{2}{\phi_h} \, \delta \phi_{inh}(0,\tau)~.
\label{omega_jump}
\end{equation}
The continuity and jump condition allow us to fix two of the three integration constants $\{\alpha_q,\beta_q,\gamma_q \}$. Finally, requiring that $\delta\omega_+(\rho,\tau)$ is fast-falling as one approaches $\rho = \pi/2$ gives us a third condition, $\gamma_q = e^{\pi |q|} \beta_q$, which in turn allows us to completely determine the linearised profile $\delta \omega(\rho,\tau)$. One finds:
\begin{eqnarray}\label{betaq}
\beta_q  =  i \frac{\kappa}{\phi_h} \text{sgn} \, q \, \frac{ e^{-\pi|q|}}{(1+q^2)(1-q^2)}  T_{\rho\tau} (0,q) \,. \label{bq}
\end{eqnarray}
 

Given $\delta\omega(\rho,\tau)$ we can obtain $\delta\phi(\rho,\tau)$ by evaluating the integral in (\ref{integral}). Notice that $\delta\phi(\rho,\tau)$ has no jump in its first derivative. It is also interesting to obtain an equation for the curve $\mathcal{S}$ separating the region of positive and negative $\phi$. To linear order, this reads
\begin{equation}\label{bulkcurve}
\bar{\phi}(\rho) + \delta \phi_{inh}(\rho,\tau) = 0~.
\end{equation} 
Solutions to the above equation $\rho_{\mathcal{S}}(\tau)$ will produce a curve $\mathcal{S}$ separating positive and negative values of $\phi$, and hence regions of positive and negative curvature. In the absence of matter, this curve is a circle. Given that there are no positivity constraints on $\delta \phi_{inh}(\rho,\tau)$ this original circle can become deformed in both directions, as we shall shortly explore in a concrete example. 

Once a linearised solution is found, one must impose the appropriate boundary conditions. 
For instance, let us consider the Dirichlet problem with boundary values $\{ h,\phi_b \}$ which we take to be $u$-independent with $\phi_b$ large and positive. In the absence of matter, the boundary value problem is solved by finding a $\rho_c$ such that $h = e^{2\omega(\rho_c)}$ and $\phi_b = -\phi_h \tan \rho_c$. Thus, in the absence of matter the boundary curve is fixed to by the circle $\mathcal{C} = \{u , \rho_c \}$. Upon turning on a linear perturbation, $\mathcal{C}$ will be slightly deformed to a new curve pataremeterised by $\mathcal{C}_p = \{u + \delta\tau(u), \rho_c + \delta\rho(u) \}$, where $\delta\tau(u)$ and $\delta\rho(u)$ are fixed by solving:
\begin{eqnarray}
h &=& e^{2\bar{\omega}(\rho_c)} \left( 1+ 2 \partial_u \delta\tau(u) +  2\delta \omega(\rho_c,u) + 2\partial_\rho\bar{\omega}(\rho_c)\delta\rho(u) \right)~, \label{bdy1}\\
\phi_b &=& \bar{\phi}(\rho_c) + \partial_\rho \bar{\phi}(\rho_c) \delta\rho(u) + \delta \phi_{inh}(\rho_c,u)~. \label{bdy2}
\end{eqnarray}
The above equations are solved by:
\begin{equation}\label{Csol}
\delta\rho(u) = -\frac{\delta \phi_{inh}(\rho_c,u)}{\partial_\rho \bar{\phi}(\rho_c)}~, \quad\quad \partial_u \delta\tau(u) = \frac{\partial_\rho \bar{\omega}(\rho_c)}{\partial_\rho \bar{\phi}(\rho_c)}  \delta \phi_{inh}(\rho_c,u)~.
\end{equation}
From the above we see that $\delta\tau(u)$ and $\delta\rho(u)$ are indeed $\mathcal{O}(\kappa)$ quantities, justifying our assumption that the resulting curve lies near the unperturbed curve $\mathcal{C} = \{ u,\rho_c\}$.

\subsection{Example}\label{example}

\begin{figure}[h!]
        \centering
        \subfigure[]{
                \includegraphics[scale=0.4]{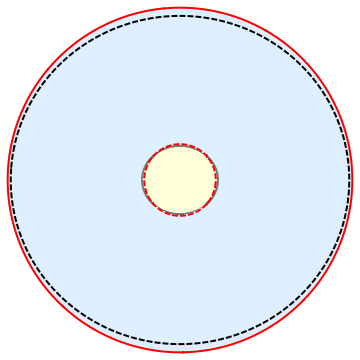}} \quad\quad\quad\quad\quad\quad
         \subfigure[]{
                \includegraphics[scale=0.4]{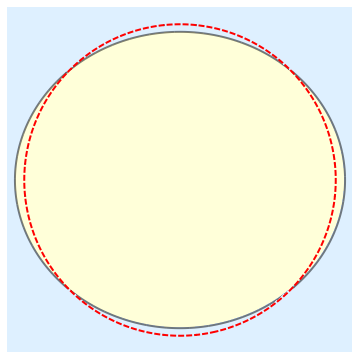}}
                 \caption{{\footnotesize The perturbative solution for $m=-\phi_h=1$. (a) shows the full perturbative solution on the unit disk, while (b) zooms in the interpolating region. The unperturbed interpolating geometry is given by a separation between negative (blue) and positive (yellow) curvature at $\rho=0$ (dashed, red circle). The perturbation for {{$\kappa h_1^2 = 2.3$}}, generates the new curve with $\phi=0$ that is given by the black line in the plot. The black dashed line close to the boundary in Fig. (a) shows boundary conditions given by $\phi_b=30$, $h=900$.}}
\label{fig_mone}
\end{figure}

As a concrete example, suppose the matter field is characterised by a single harmonic $m\ge1$ and is real valued. The solution for $\chi(\rho,\tau)$ is simply:
\begin{equation}
\chi (\rho,\tau) = 2 \, h_{m} \, e^{m (\rho-\pi/2)} \cos(m \tau) \,,
\end{equation}
from which it immediately follows that:
\begin{equation}
T_{\rho\tau}^\chi (\rho, \tau) = -2 h_m^2 m^2 e^{2 m (\rho-\pi/2)} \sin (2 m \tau)~.
\end{equation}
It is then straightforward to obtain the perturbative solution. We first evaluate the integral in equation (\ref{integral}) for the given stress tensor. 
The next step is to find the curves where the dilaton vanishes. This can be done numerically and shown for an example with $m=1$ in figure \ref{fig_mone}. 
We still need to fix the boundary conditions. Given $h$ and $\phi_b$, then $\delta \rho(u)$ and $\delta\tau(u)$ are determined by (\ref{Csol}). In figure \ref{fig_mone} we display the solution with $h=900$, $\phi_b=30$. 

\begin{figure}[h!]
        \centering
        \subfigure[$\kappa h_1^2=8.3$]{
                \includegraphics[scale=0.4]{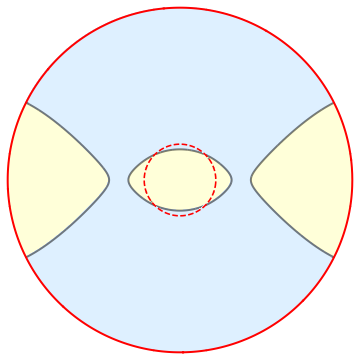}}  \quad\quad\quad\quad\quad\quad
         \subfigure[$\kappa h_1^2=8.8$]{
                \includegraphics[scale=0.4]{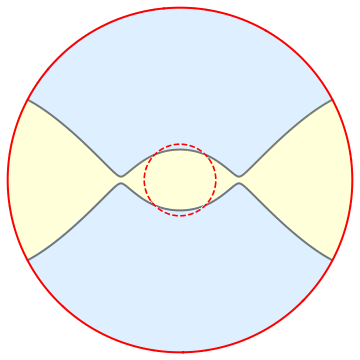}}
                 \caption{{\footnotesize Perturbative solution with $m=-\phi_h=1$ on the unit disk. Yellow regions indicate positive curvature while blue ones, negative. The interpolating region with $\phi=0$ is the solid curve between them. The red dashed circle shows the interpolating region of the unperturbed geometry. As the strength of the perturbation increases, the positive curvature regions grow until negative curvature regions become disconnected.}}
\label{fig_mone2}
\end{figure}

More generally, if we take $\partial\mathcal{M}$ to lie at $\rho_c \approx \pi/2$ and $\tau=u$, the condition that $\phi_b$ remains positive and large along $\partial \mathcal{M}$ enforces:
\begin{equation}\label{example_phih}
|\phi_h| \gtrsim \frac{\kappa \,  h_m^2 \, m}{(4m^2+1)} \left(1 +  \frac{8m^2e^{-2\pi m}}{(4m^2-1)} \right)~.
\end{equation}
Relaxing the above condition leads to solutions where regions of negative and positive curvature might separate, as shown in figure \ref{fig_mone2}. If we impose a Dirichlet boundary condition for which $\phi_b$ is everywhere positive along $\partial \mathcal{M}$, solutions such as the one in figure \ref{fig_mone2} (b) are no longer allowed. Perhaps this suggests for the Dirichlet problem with AdS$_2$ asymptotia, one can entirely remove the dS$_2$ region in the interior by turning on strong enough matter sources at $\partial \mathcal{M}$. It would be interesting to futher explore this at the non-linearised level.

\subsection{Remarks on the non-linear problem}

We would like to end this section with some remarks on the general non-linear problem. Again, we will consider the limit where there is a sharp transition from positive to negative curvature. The geometric problem consists of gluing a constant negative curvature geometry to a constant positive curvature geometry across some curve $\mathcal{S}$. In the conformal gauge, constant curvature metrics in two-dimensions are given by solutions to the Liouville equation. To express the space of solutions, it is convenient to introduce a complex coordinate $z = e^{\rho+ i\tau}$. A large class of positive constant curvature geometries are:
\begin{equation}
ds^2 = \frac{4 f_+'(z) g_+'(\bar{z})}{\left(1+f_+(z) {g_+(\bar{z})} \right)^2}  \, dz  \, d\bar{z}~,
\end{equation}
where $f_+(z)$ and $g_+(\bar{z})$ are meromorphic and anti-meromorphic functions giving rise to a real geometry. For example the round metric on the two-sphere has $f_+(z)=z$ and $g_+(\bar{z})=\bar{z}$. Negative constant curvature geometries are:
\begin{equation}
ds^2 = \frac{4 f_-'(z) g_-'(\bar{z})}{\left(1-f_-(z) {g_-(\bar{z})} \right)^2}  \, dz \,  d\bar{z}~.
\end{equation}
The standard hyperbolic geometry has $f_-(z)=z$ and $g_-(\bar{z})=\bar{z}$. The hyperbolic cylinder is given by $f_-(z) = e^{i\log z}$ and $g_-(\bar{z}) = e^{i\log \bar{z}}$.  Poles in $f_-(z)$ and $g_-(\bar{z})$ translate to conical defects in the two-dimensional geometry.

We wish to glue the two geometries together along an arbitrary closed curve $\mathcal{S}$, such that both the metric and extrinsic curvature matches along the curve. The equation for $\mathcal{S}$ will be some non-holomorphic function ${s}(z,\bar{z})=0$. Thus, $f_+(z)$ and $g_+(\bar{z})$ should be analytic in $\mathcal{U}_{\mathcal{S}}$.  Throughout the region $\mathcal{U}_{\mathcal{S}}$ within $\mathcal{S}$ we require a smooth constant positive curvature geometry. By the Riemann mapping theorem, we can map $\mathcal{U}_{\mathcal{S}}$ to the open disk with circular boundary. We can then fill the interior of the disk with the standard metric on the two-sphere and glue it to the appropriate hyperbolic geometry. Finding the explicit form of the map is generally complicated. Upon performing the map, the boundary values of the fields will transform. It is important to understand how the boundary values of the fields transform since they are to be interpreted as sources for the operators in a putative dual quantum mechanics.

\section{Boundary soft mode} \label{schwartzian}

In this section we consider the effective theory of the soft mode arising close to the AdS$_2$ boundary. We emphasise that the AdS$_2$ clock near the boundary of the unperturbed, or slightly perturbed, interpolating geometry is the isometric direction of the hyperbolic cylinder. As mentioned earlier, this results in a sign difference for the boundary action (\ref{bdyaction}) as compared to that of the hyperbolic disk. 

\subsection{Soft action}

Let us take $h$ and $\phi_b$ to be $u$-independent, with $\phi_b$ positive and the two scaling as $\{\sqrt{h},\phi_b\} = \Lambda \{1,\tilde{\phi}_b \}$ in the large-$\Lambda$ limit. We take the proper size of the boundary circle to be $\Lambda \, \tilde{\beta}$ such that $u \sim u + \tilde{\beta}$. The boundary action for the interpolating geometry (\ref{bmetric}) is found by calculating the extrinsic curvature for the curve $\mathcal{C} = \{\tau(u),\rho(u) \}$. 

If we remain off-shell by leaving $\tau(u)$ unfixed, we can calculate the boundary action for the interpolating geometry:
\begin{eqnarray}\label{gbdy}
S_{bdy} = \frac{\tilde{\phi}_b}{\kappa} \int du \left( \frac{1}{2} \left(\partial_u \tau(u)\right)^2 - \text{Sch} \left[ \tau(u),u \right] \right) \,.
\end{eqnarray}
In deriving the above action, we have related $\rho(u)$ to $\tau(u)$ via (\ref{induced}):
\begin{equation}
\Lambda^2 = \frac{(\partial_u \rho(u))^2 + (\partial_u \tau(u))^2 }{\cos^2\rho(u)}~.
\end{equation}
In the large $\Lambda$-limit, we have $\Lambda \cos\rho(u) \approx \partial_u \tau(u)$ such that $\rho(u)$ must be parameterically near $\rho = \pi/2$. This forces the boundary curve $\mathcal{C}$ to live parametrically close to the AdS$_2$ boundary. For $\tau(u)$ non-compact, the theory (\ref{gbdy}) is invariant under the $SL(2,\mathbb{R})$ transformation:
\begin{equation}
\sigma(u) \to \frac{a \, \sigma(u) + b}{c \, \sigma(u) + d}~, \quad\quad a d - b c = 1~,
\end{equation}
with $a,b,c,d$ real, and $\sigma(u) = \tanh  \tau(u)/2$. Due to the fact that $\tau(u)$ is compact, the $SL(2,\mathbb{R})$ invariance is broken to a $U(1)$ subgroup corresponding to shifts in $\tau(u)$. 

The equations of motion stemming from (\ref{gbdy}) are:
\begin{equation}\label{eomS}
\partial_u^2 \tau = \partial_u \left( \frac{1}{\partial_u\tau} \partial_u \left( \frac{\partial_u^2 \tau }{\partial_u \tau} \right) \right)~.
\end{equation}
One solution to the above equations is $\tau(u) = 2\pi u /\tilde{\beta}$, such that $u \sim u + \tilde{\beta}$. In \cite{Maldacena:2016upp, Jensen:2016pah}, $\tilde{\beta}$ is interpreted as the temperature of a putative ultraviolet system. An example where this occurs is the SYK model, for which $u$ becomes the clock of the dual quantum mechanical fermions. We will keep $\tilde{\beta}$ as a parameter, and occasionally interpret it as a temperature. In the case where our bulk theory has an additional scale, such as the field range of the interpolating region in the dilaton potential, the theory already contains a bulk tunable parameter that can be viewed as a temperature. This is in sharp contrast to linear dilaton potentials for which the bulk geometry is entirely fixed at the classical level. 
 
The on-shell boundary action for the $\tau(u) = 2\pi u /\tilde{\beta}$ saddle becomes:
\begin{equation}\label{Sbdylinear}
S_{bdy,cl} =  -\text{sign}  \phi_h \frac{\tilde{\phi}_b}{\tilde{\beta} \kappa}~.
\end{equation} 
If we interpret $\tilde{\beta}$ as an inverse temperature, we see that (\ref{Sbdylinear}) is linear in the temperature.
For $\kappa>0$ the specific heat with respect to variations of $\tilde{\beta}$ is negative for the interpolating solution, which as we recall has $\phi_h<0$.\footnote{Note that for a class of generalised potentials that is analysed in appendix \ref{gammaSch}, it is possible to obtain interpolating geometries with $\kappa>0$ and a positive specific heat.} Nevertheless, as we shall soon see, the $\kappa>0$ thermal saddle is locally stable with respect to small variations of $\tau(u)$.

For the remainder of the solution space, we can map (\ref{eomS}) to the equations of motion of Liouville quantum mechanics \cite{Bagrets:2016cdf}:
\begin{equation}\label{LQM}
\partial_u^2 \xi(u) = e^{2\xi(u)}~,
\end{equation} 
with $\partial_u\tau(u) = e^{\xi(u)}$. The solutions to (\ref{LQM}) are:
\begin{equation}\label{ZZ}
e^{2\xi(u)} = a^2 \, \text{sec}^2 a u~.
\end{equation}
Due to the condition $\xi(u) = \xi(u+\tilde{\beta})$, we require $a = n \pi/\tilde{\beta}$ with $n\in \mathbb{Z}$. All these saddles have a divergence at certain values of $u$. Evaluating the on-shell action for the solutions (\ref{ZZ}) reveals a divergent answer.
The action in terms of $\xi$ is given by:
\begin{equation}
S_{\xi} = \frac{\tilde{\phi}_b}{\kappa} \int du \left[ \left(\partial_u \xi\right)^2 + \frac{1}{2}e^{2\xi(u)} \right]~.
\end{equation}
As shown in  \cite{Bagrets:2016cdf}, the path integral measure is flat in $\xi(u)$ with the zero mode removed.

\subsubsection*{{\it Case (i): $\kappa>0$}}

For $\kappa > 0$ the dominant saddle has $\phi_h>0$ such that the two-dimensional geometry is the hyperbolic disk. The interpolating solution is a sub-leading contribution to the thermal partition function. 
Picking $\tilde{\beta}=2\pi$, and expanding the soft action around the saddle as $\tau(u) = u+ \delta\tau(u)$ we obtain the following perturbative action for the Schwarzian piece:
\begin{equation} \label{sch_global}
S_{fluct} = \frac{\tilde{\phi}_b}{2\kappa} \int du \, \left( \left(\partial^2_u \delta\tau(u) \right)^2 + \left(\partial_u \delta\tau(u) \right)^2  \right)~.
\end{equation}
Interestingly, though the interpolating geometry is a sub-dominant saddle, the action for fluctuations (\ref{sch_global}) of its soft boundary mode is Gaussian suppressed, allowing for a perturbative analysis of $\delta\tau(u)$ in the sub-dominant saddle. 
We will analyse this shortly. 

\subsubsection*{{\it Case (ii): $\kappa<0$}}

We now consider $\kappa<0$. In this case, the boundary action for the interpolating geometry (\ref{gbdy}) dominates over that of the hyperbolic disk. This is consistent with the analysis of \cite{Anninos:2017hhn}. However, care must be taken with the perturbative analysis of fluctuations. Indeed, for $\kappa<0$, the fluctuation action (\ref{sch_global}) is Gaussian unsuppressed. One possibility is that we must consider a different ensemble for $\kappa<0$. For instance, we could consider a mixed Neumann-Dirichlet ensemble where $\pi_\phi = -K$ and $h$ are fixed, rather than the standard Dirichlet problem we have been considering so far. Fixing the extrinsic curvature in Euclidean quantum gravity has been considered in other contexts also (see for example \cite{Hartle:1983ai,Witten:2018lgb}). In going to this ensemble, we must allow $\phi_b(u)$ to fluctuate, and hence, it is no longer guaranteed that our geometry will be asymptotically AdS$_2$.
Another possibility is that we must view the $\kappa<0$ theory as part of a larger theory with non-negative $\kappa$. One possible lesson is that for $\kappa<0$, we should no longer interpret $\tilde{\beta}$ as a temperature. 

An example of a well defined theory where one encounters a term in the action corresponding to a Schwarzian action with negative coefficient arises if one considers coupling the soft mode action to a non-trivial geometry $\tilde h(u)$. In order to covariantise this action it is useful to write:
\be
\text{Sch}[\tau(u),u] = F[A,u] \equiv \partial_u A_u - \frac{1}{2} A_u A_u~, \quad\quad A= \partial_u \log  \partial_u \tau(u)  \, du~.
\ee
Notice that $A$ transforms as a connection under changes of coordinates $u \rightarrow f(u)$.\footnote{Incidentally, $A$ also transforms as a gauge connection for the conformal symmetry action on target space $\tau \rightarrow \sigma(\tau)$ and $F$ is a weight 2 tensor invariant under the global $SL(2,\mathbb{R})$ sub-algebra of this local group.} Provided an affine connection $\Gamma$, it is now easy to covariantise this action in terms of $\mathcal{A} = A - \Gamma$. In a manifold with a metric $\tilde h(u)$, $\Gamma$ is given by
\be\label{affine}
\Gamma =\tilde h^{-1/2} \partial_u \tilde{h}^{1/2} \, du~.
\ee
The associated covariant Schwarzian action becomes:
\be
\text{Sch}_{\tilde h}[\tau(u),u] = \tilde h^{-1} \mathcal{F}[\mathcal{A}, u] \equiv  \tilde h^{-1} \left[ \mathcal{D}_u \mathcal{A}_u - \frac{1}{2} \mathcal{A}_u \mathcal{A}_u \right]
\ee
\noindent where $\mathcal{D}_u \equiv \partial_u - \Gamma_u$ is the usual covariant derivative and we have included an inverse metric $\tilde h^{-1}$ to make the Schwarzian a scalar density. Now something interesting happens as we expand out this expression:
\be
\mathcal{F}[\mathcal{A}, u] =  \partial_u A_u - \frac{1}{2} A_u A_u -  \partial_u \Gamma_u + \frac{1}{2} \Gamma_u \Gamma_u = F[A,u] - F[\Gamma,u]
\ee
Putting all this together we can write a fully covariant action as:
\be
S[\tau, \Gamma,\tilde h] = -\frac{1}{\kappa} \int du \sqrt{\tilde h} \tilde \phi_b \, \tilde h^{-1} \left( F[A,u] - F[\Gamma,u] \right)~.
\label{diegoformula}
\ee
It is clear from the above expression that regardless of the original sign of $\kappa$ we obtain the difference of two identical actions with different signs. These can of course differ on the actual variables of integration in the path integral. As expected in first order formalisms of gravity, if we take $\Gamma$ to be an independent variable we obtain (\ref{affine}) as the equation of motion. We will elaborate on these issues in future work. 

\section{Matter perturbations}
\label{sec_otoc}

We now consider the effect of a massless free scalar, with boundary value $\chi(u)$, to the physics of the soft mode. The contribution to on-shell action from the matter theory (\ref{matterS}) is given by:
\begin{equation}\label{softmatter}
S_{\chi} = -\frac{1}{2} \int d u_1 d u_2 \left( \frac{\tau'(u_1) \tau'(u_2) }{\sin^2 \frac{\tau(u_1)-\tau(u_2)}{2}} \right) \bar{\chi}(\tau(u_1)) {\chi}(\tau(u_2))~.
\end{equation} 
In the above expression, we have assumed that the boundary curve $\mathcal{C} = \{ \tau(u),\rho(u)\}$ always remains near the asymptotic AdS$_2$ boundary. In addition to (\ref{softmatter}), the total on-shell action also contains a contribution from the Schwarzian action (\ref{bdyaction}) and an interior contribution $S_{int}$ given in (\ref{onshell2}). 
Thus, the generating function for boundary correlations of the scalar $\chi$ contains an additional contribution when compared to the pure Jackiw-Teitelboim theory with a linear dilaton potential. There, the geometry is always pure AdS$_2$ and the whole problem can be mapped to a calculation at the boundary \cite{Maldacena:2016upp,Jensen:2016pah,Engel}. Here, the soft boundary physics carries an imprint from the bulk interpolating region. 
This can be seen rather directly from our perturbative equations (\ref{integral}) and (\ref{Csol}). The contribution to (\ref{Csol}) coming from the integral over the stress-tensor is equivalent to the equation of motion stemming from the boundary theory (\ref{gbdy}) plus (\ref{softmatter}). However, there is also a contribution to (\ref{Csol}) from the second integral in (\ref{integral}). This term is due to the interpolating region. 

It is instructive to obtain an equation for the off-shell curve parameter $\tau(u) \approx u + \delta\tau(u)$. This can be done by considering the ADM mass of the theory \cite{Grumiller:2007ju} along the curve $\mathcal{C}$:
\begin{equation}\label{ADMmass}
M_{ADM} = \sqrt{h} \left(-n^a \partial_a \phi + \phi_b \right)~.
\end{equation}
We can evaluate $M_{ADM}$ either using the $T_{\rho\rho}$ or $T_{\rho\tau}$ equations of motion. Given a slightly perturbed metric,
\begin{equation}\label{pertmetric}
ds^2 = e^{2 \bar{\omega}(\rho)}\left(1+2\delta\omega(\rho,\tau) \right)(d\rho^2+d\tau^2) \,,
\end{equation}
we evaluate (\ref{ADMmass}) along the curve $\mathcal{C} = \{u+\delta\tau(u),\rho_c + \delta \rho(u) \}$, where $\delta\rho(u)$ is expressed in terms of $\delta\tau(u)$ and $\delta\omega(\rho_c,u)$ by fixing the induced metric $h = e^{2\bar{\omega}(\rho_c)}$ of (\ref{pertmetric}). Working to linear order in $\delta\tau(u)$ and $\delta\omega(\rho_c,u)$, and using that $\phi_b = -{\phi}_h \tan\rho_c$, we obtain the following expression:
\begin{equation}\label{bdyEOM}
\tilde{\phi}_b \left(\partial_u^4 -\partial_u^2 \right) \delta\tau(u) =  - \, \kappa \,T_{\tau\rho}(\pi/2,u) + 3\, h \, \tilde{\phi}_b  \, \partial_u \delta\omega(\rho_c,u)~.
\end{equation}
We also work in the limit $\rho_c \to \pi/2$, which is well defined so long as $\delta\omega(\rho_c,u)$ goes as $\mathcal{O}((\rho_c-\pi/2)^2)$ as we approach $\rho_c=\pi/2$. The left hand side of (\ref{bdyEOM}) is given by linearising the equation of motion of the Schwarzian theory (\ref{gbdy}). The first term on the right hand side of (\ref{bdyEOM}) comes from varying (\ref{softmatter}) with respect to $\tau(u)$. The last term in (\ref{bdyEOM}) is due to the slightly perturbed Weyl factor and encodes the information of the interior curve $\mathcal{S}$. Upon inserting the on-shell value for $\delta\omega(\rho,\tau)$, the solutions of (\ref{bdyEOM}) are the same as our linear solutions (\ref{Csol}). Since $\delta\omega(\rho,\tau)$ is fixed by $T_{\rho\tau}(\rho,\tau)$ on-shell, we end up an equation for $\delta\tau(u)$ that is completely determined by the boundary data $\chi(u)$. Recall that in momentum space, the linearised solution for $\delta\omega$ near the boundary can be obtained by expanding (\ref{deltaomegaplus}) near $\rho=\pi/2$:
\begin{equation}
\delta\tilde{\omega}_q \equiv \lim_{\rho\to\pi/2} \frac{\delta\omega_q(\rho)}{(\rho-\pi/2)^2} =  -\frac{2}{3}i \, \frac{\kappa}{\tilde{\phi}_b} \frac{q \, e^{-\frac{\pi  \left| q\right| }{2}}}{(1-q^2)} T_{\rho\tau }(0,q) ~.
\end{equation}
In position space, this becomes a convolution:
\begin{equation}\label{convG}
\delta\tilde{\omega}(u) = -\frac{2}{3} \, \frac{\kappa}{\tilde{\phi}_b} \partial_u \int dw\, G(u-w)\,T_{\rho\tau}(0,w)~, \quad\quad G(u) \equiv \sum_{q \neq \{\pm1,0\}} \frac{e^{-\frac{\pi|q|}{2}}}{(1-q^2)} e^{i q u}~.
\end{equation}
Explicitly:
\begin{eqnarray}\label{Gtrig}
G(u) = \frac{1}{2}\, \mathfrak{Re} \left[\sinh \left(\frac{\pi}{2}- i u\right) -\cosh \left(\frac{\pi}{2} - i u\right)-4 \log \left(1-e^{-\frac{\pi }{2}+i u}\right) \sinh \left(\frac{\pi}{2} - i u\right)-2\right]~. 
\end{eqnarray}
Notice that $G(u)$ is an oscillatory function in $u$.

\subsection{Boundary action}

It is of interest to calculate the boundary four-point function for the matter fields which we express as $\chi=(\chi_1+i\chi_2)/\sqrt{2}$, with $\chi_1$ and $\chi_2$ real. We will set $\tilde{\beta} = 2\pi$ unless otherwise specified. At tree-level, we must compute the on-shell Euclidean action for the linearized perturbations $\delta \tau$ and $\chi$. It is given by:
\begin{equation}\label{chibdy}
S_{bdy}[\delta\tau(u),\chi(u)] = S_{fluct}[\delta\tau(u) ] - \int du \left(T_{\rho\tau}(\pi/2,u)-\frac{3\tilde{\phi}_b}{\kappa} \partial_u \delta\tilde{\omega}(u) \right)  \delta\tau(u)~,
\end{equation}
with $S_{fluct}[\delta\tau(u)]$ containing the fluctuations of the Schwarzian given in (\ref{sch_global}) and $\delta\tilde{\omega}(u)$ given in (\ref{convG}). Since $T_{\rho\tau}$ and $\delta\tilde{\omega}$ are quadratic in $\chi_i(u)$, we can read off a cubic interaction between $\chi_i(u)$ and $\delta \tau(u)$ directly from (\ref{chibdy}). Moreover, given that:
\begin{equation}
T_{\rho\tau}(\pi/2,u) = -\frac{1}{4}  \left( \partial_u \chi(u) \int dw \frac{\bar{\chi}(w)}{\sin^2 \frac{u-w}{2}} + h.c. \right)~,
\end{equation}
it follows that (\ref{chibdy}) is non-local in $u$, when viewed as a functional of $\chi(u)$ and $\delta\tau(u)$. Similarly, 
\begin{equation}\label{chitilde}
T_{\rho\tau}(0,u) = -\frac{1}{4}  \left( \partial_u \tilde{\chi}(u) \int dw \frac{\bar{\tilde{\chi}}(w)}{\sin^2 \frac{u-w}{2}} + h.c. \right)~,  
\end{equation}
where we have further defined:
\begin{equation}
\tilde{\chi}(u) \equiv \int d w  \frac{\sinh \pi/2  \, {\chi(w)}}{\cosh\pi/2- \cos(u-w)}~.
\end{equation}
The reason for taking a convolution of the boundary matter source $\chi(u)$ is that there is a relative factor of $e^{-|q|\pi/2}$ for the Fourier modes of $\chi(\rho,\tau)$ when comparing $\rho=0$ to $\rho=\pi/2$. The convolution suppresses the small scale structure of $\chi(u)$. As a result we can rewrite the second term in (\ref{chibdy}) as:
\begin{equation}
 -\frac{1}{8}\int \frac{du_1 du_2}{\sin^2 \frac{u_1-u_2}{2}} \left[ {\chi(u_1) \bar{\chi}(u_2)} \mathcal{B}(u_1,u_2) + \tilde{\chi}(u_1) \bar{\tilde{\chi}}(u_2) \mathcal{E}(u_1,u_2) + h.c. \right]~,
\end{equation}
where we have defined:
\begin{eqnarray}
\mathcal{E}(u_1,u_2) &\equiv&  \left( \partial_{u_1} \delta\mathcal{T}(u_1)+ \partial_{u_2} \delta \mathcal{T}(u_2)  -  \frac{\delta\mathcal{T}(u_1) -\delta \mathcal{T}(u_2)}{\tan \frac{u_1-u_2}{2}}  \right)~, \\
\mathcal{B}(u_1,u_2) &\equiv& \left( {\partial_{u_1}\delta \tau(u_1)+ \partial_{u_2} \delta \tau(u_2) } -  \frac{\delta \tau(u_1) - \delta\tau(u_2) }{\tan \frac{u_1-u_2}{2}}   \right)~,
\end{eqnarray}
and $\delta\mathcal{T}(w) \equiv 2 \int du  \, G(w-u) \partial^2_{u} \delta\tau(u)$, with $G(u)$ given in (\ref{convG}).

\begin{figure}[h!]
\begin{center}
 \includegraphics[scale=0.6]{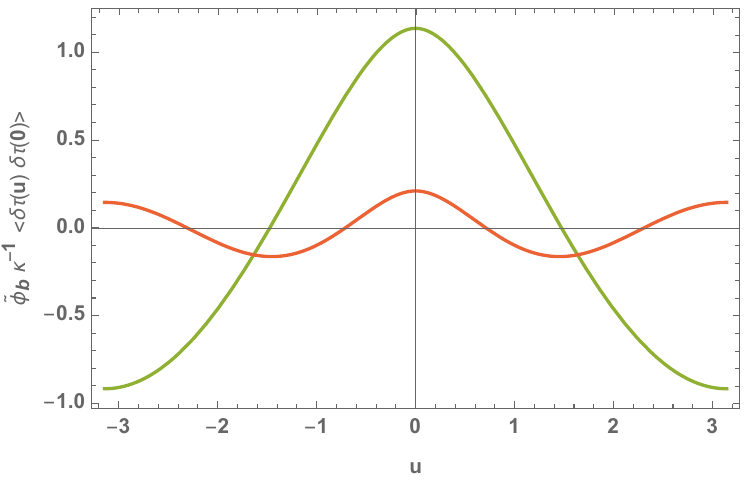}
\end{center}
\vspace{-0.8cm}
  \caption{{\footnotesize The propagator in equation (\ref{propagator_global}) as a function of $u$ is shown in green. In red, we plot the propagator for the perturbative Schwarzian action resulting near the boundary of the hyperbolic disk.}} \label{fig_props}
\end{figure}

From the action of small soft mode fluctuations (\ref{sch_global}) with $\kappa>0$, we can extract the propagator of the mode $\delta\tau(u)$ which in Fourier space reads:
\begin{equation}
\langle \delta\tau_m \delta\tau_{n} \rangle = \frac{\kappa}{\tilde{\phi}_b} \, \frac{\delta_{m+n}}{m^4+m^2}~.
\end{equation}
Notice that $m=0$ is a zero mode that must be excised from the configuration space. Going back to position space, the propagator becomes:
\begin{equation}
\langle \delta\tau (u) \delta\tau (0) \rangle = \frac{\kappa}{\tilde{\phi}_b} \left( \frac{ (\left| u\right|- \pi) ^2 }{2} - \frac{\pi}{\sinh \pi}  \cosh (\left| u\right| - \pi ) +1 -\frac{\pi ^2}{6} \right) ~. \label{propagator_global}
\end{equation}
We display the $\delta\tau(u)$ propagator in figure \ref{fig_props}. This result is valid for $u  \in [-\pi,\pi)$ and periodically continued for other $u$.
To obtain the contribution to the $\chi_i$ four-point function from (\ref{chibdy}), we must integrate out $\delta\tau(u)$. To leading order, this is given by calculating the tree-level exchange diagram of $\delta\tau(u)$. 

\subsection{Four-point functions}

We have now collected all the ingredients necessary to calculate the tree-level, connected, four-point function. It is given by:
\begin{equation}
{\langle \chi_1(u_1) {\chi}_1(u_2) \chi_2(u_3) {\chi}_2(u_4) \rangle_c} = \frac{1}{16} \left< \left(\frac{\mathcal{B}(u_1,u_2)}{\sin^2 \frac{u_1-u_2}{2}} + \tilde{\mathcal{E}}(u_1,u_2) \right)\left(\frac{\mathcal{B}(u_3,u_4)}{\sin^2 \frac{u_3-u_4}{2}} + \tilde{\mathcal{E}}(u_3,u_4) \right) \right> \,,
\end{equation}
where we have defined:
\begin{equation}\label{Etilde}
\tilde{\mathcal{E}}(u_1,u_2) \equiv \int dw_1 dw_2 \, \frac{\sinh\pi/2}{\cosh\pi/2 - \cos(u_1-w_1)} \, \frac{\mathcal{E}(w_1,w_2)}{\sin^2 \frac{w_1-w_2}{2}} \, \frac{\sinh\pi/2}{\cosh\pi/2 - \cos(w_2-u_2)}~.
\end{equation}

The calculations are somewhat lengthy, but straightforward. In appendix \ref{app_otoc} we give some details of the calculations. The time ordering $u_4<u_2<u_3<u_1$ is the one relevant for out-of-time-ordered correlations \cite{larkin} upon suitable analytic continuation. This can be done in many different ways. A particularly simple configuration is given by placing the four operators at equally-spaced points along the thermal circle, and then evolving only the two diametrically opposed ones along the real-time axis:
\begin{eqnarray}
F(t) \equiv  {\left\langle \chi_1( {\pi}/{2}) \chi_2(  i t)  {\chi}_1(-{\pi}/{2}) {\chi}_2( -\pi + i t) \right\rangle_c} \label{ef_of_t} = \text{Tr} \hat{\rho} \left[  \hat{y}  \, \hat{\chi}_1 \, \hat{y}  \, \hat{\chi}_2(t) \, \hat{y}  \, \hat{\chi}_1 \, \hat{y}  \, \hat{\chi}_2(t) \right]~,
\end{eqnarray}
where $\hat{\rho} \equiv e^{-\tilde{\beta}\hat{H}}$ and $\hat{y} \equiv \hat{\rho}^{1/4}$. It has been argued that quantum systems in thermal equilibrium with a large number of degrees of freedom, $N$, and a parametrically large separation between dissipation and scrambling time obey \cite{Maldacena:2015waa}:
\begin{equation}\label{Lyapunov}
F(t) = f_0 - \frac{f_1}{N} \exp (\lambda_L t) \,, ~ ~ \text{with} ~ ~ \lambda_L \leq \frac{2\pi}{\tilde{\beta}} \,.
\end{equation}
Here, $\lambda_L$ is the Lyapunov exponent, $f_0$ and $f_1$ are positive order-one constants (independent of $N$, $\tilde{\beta}$, $t$), and $\tilde{\beta} \ll t \ll \tilde{\beta}\log N$. It turns out that black holes in Einstein gravity saturate this bound. The AdS$_2$ black hole with running dilaton also saturates the chaos bound. 

We are interested in the four-point function $F(t)$ for the interpolating geometry. The important point is that the $\delta\tau(u)$ propagator (\ref{propagator_global}) is built from hyperbolic functions, which upon analytic continuation become oscillatory. This already implies that the piece of $F(t)$ stemming purely from the $\mathcal{B}(u,w)$ contribution will oscillate in $t$. Upon reinstating $\tilde{\beta}$, the frequency of oscillation is $2\pi/\tilde{\beta}$, which is the same as the Lyapunov coefficient. Analytically:
\begin{eqnarray}\label{nochaos}
{\left \langle  \mathcal{B}(\pi/2,-\pi/2 ) \mathcal{B}(2\pi i t /\tilde{\beta}, -\pi+ 2\pi i t/\tilde{\beta}) \right \rangle_c} = \frac{2\kappa \tilde{\beta}}{\tilde{\phi}_b}
\left( \pi   \text{csch}\frac{\pi }{2}  \cos \frac{2 \pi  t}{\tilde{\beta} }-2 \right) \,.
\end{eqnarray}
The piece of $F(t)$ stemming from the $\mathcal{E}(u,w)$ contribution contains trigonometric functions both in (\ref{Gtrig}), as well as in the convolution (\ref{chitilde}). Recalling that we must calculate correlators of $\tilde{\mathcal{E}}(u,w)$ in (\ref{Etilde}), the only piece that is subject to analytic continuation is the $\cos(u-w)$ in the denominator of (\ref{Etilde}). All the relevant integrals appearing in the convolutions are finite and well defined. Consequently, upon setting $u_1 = \pi/2$, $u_2 = -\pi/2$, $u_3 = i t$ and $u_4 = -\pi + i t$, the contribution from $\tilde{\mathcal{E}}(u,v)$ will either oscillate or decay exponentially at large $t$.  In appendix \ref{app_otoc} we provide numerical evidence of this behaviour. 

In conclusion, the boundary out-of-time-ordered four-point function of $\chi(u)$ for the interpolating geometry does not display the exponential Lyapunov behaviour (\ref{Lyapunov}) observed for black holes in AdS. Instead, we observe oscillatory behaviour in Lorentzian time. This is an interesting distinction between de Sitter-like and black-hole-like horizons whose microscopic interpretation we will discuss in future work.

%
%

\section{Lorentzian deformations}\label{shocksec}

Having discussed several aspects of the Euclidean problem, we move on to some features of perturbations in Lorentzian signature. Specifically, we will consider the effect of a null energetic pulse travelling from the AdS$_2$ boundary through the dS$_2$ horizon. 

\subsection{Lorentzian background}

It is convenient to work in light-cone coordinates, such that our gauge-fixed background is given by:
\begin{equation}
ds^2 = e^{2\omega(x^+,\, x^-)} dx^+ dx^-~.
\end{equation}
The relation to the Schwarzschild like coordinates (\ref{schwcoord}) in the $x^+>0$, $x^->0$ quadrant is:
\begin{equation}
\pm \log x^{\pm} = t \pm \int^\rho \frac{dr}{{N(r)}}~.
\end{equation}
The Ricci scalar is given by $R = - 8 e^{-2\omega} \partial_+\partial_-\omega$. The non-vanishing Christoffel symbols are:
\begin{equation}
\Gamma^+_{++} = 2\partial_+ \omega~, \quad \Gamma^-_{--} = 2 \partial_- \omega~.
\end{equation}
The background solution is given as follows. The dS$_2$ region is covered by the range $-1<x^+ x^- < 1$. The metric and dilaton take the form:
\begin{equation}\label{dskruskal}
e^{2\bar{\omega}(x^+,x^-)} = \frac{4}{(1+x^+ x^-)^2}~, \quad\quad \bar{\phi}(x^+,x^-) =  |\phi_h| \, \left( \frac{x^+ x^- - 1}{x^+ x^-+1} \right) ~.
\end{equation}
The dS$_2$ horizon resides at $x^+ x^-=0$. The future boundary in the interior of the dS$_2$ horizon resides at $x^+ x^- = -1$. The transition region occurs near $x^+ x^- = 1$. The AdS$_2$ region is covered by $1 < x^+ x^- < e^\pi$ and described by:
\begin{equation}\label{nullAdS2}
e^{2\bar{\omega}(x^+,x^-)} = \frac{1}{x^+ x^-}\sec^2 \frac{\log x^+ x^-}{2}~, \quad\quad \bar{\phi}(x^+,x^-) = |\phi_h| \tan \frac{\log x^+ x^-}{2}~.
\end{equation}
The AdS$_2$ boundary resides at $x^+ x^- = e^\pi$.


\subsection{Massless matter}

The matter content we consider will be a free massless scalar described by the action:
\begin{equation}
S_\chi = -\frac{1}{2} \int d^2x \sqrt{-g}g^{ab}  \partial_a \chi \partial_b \chi~.
\end{equation}
The solutions to the massless wave-equation are given by a sum of left and right moving waves:
\begin{equation}
\chi(x^+,x^-) = \chi_+(x^+) + \chi_-(x^-)~.
\end{equation}
In Lorentzian space we can turn on a purely chiral excitation. This would correspond to a complex, holomorphic solution in Euclidean space. Let us consider the case $\chi_-(x^-) = 0$.  The equations of motion are (see for example \cite{Almheiri:2014cka}):
\begin{eqnarray}
e^{2\omega(x^+,x^-)} \partial_+ e^{-2\omega(x^+,x^-)} \partial_+ \phi &=& -{\kappa} \, T^\chi_{++}(x^+)~, \\
e^{2\omega(x^+,x^-)} \partial_- e^{-2\omega(x^+,x^-)} \partial_- \phi &=& 0~, \\ 
4\partial_+ \partial_- \phi - {e^{2\omega(x^+,x^-)}} \, V(\phi) &=& 0~.
\end{eqnarray} 
Let us further consider the case where the geometry is transitioning sharply from AdS$_2$ to dS$_2$. The pulse emanates from the AdS$_2$ boundary and we can solve for $\phi$ by integrating the $++$-equation of motion from the past boundary to the interpolating region. In this region we find:
\begin{equation}\label{phiT}
\partial_+\phi(x^+,x^-) =  \partial_+\bar{\phi}(x^+,x^-) - \kappa\, e^{2\bar{\omega}(x^+,x^-)}  \, \int_{-\infty}^{x^+} d\tilde{x}_+ e^{-2\bar{\omega}(\tilde{x}_+,{x}_-)} T^\chi_{++}(\tilde{x}_+)~.
\end{equation}
We can consider the shockwave limit where $T^\chi_{++}(x^+) = \alpha \delta(x^+)$. The deviation from the background value of the dilaton is:
\begin{equation}\label{shockI}
\delta \phi(x^+,x^-) = - {\alpha}{\kappa} \, e^{-2\omega(0,x^-)} \,  \Theta(x^+) \,  \int^{x^+}_0 d u \, e^{2\omega(u,x^-)} = - {\alpha}{\kappa} \left(\frac{ x^+}{x^+ x^- + 1} \right) \, \Theta(x^+)~.
\end{equation}
It is also useful to consider the following coordinate transformation:
\begin{equation}
\tilde{x}^+ = x^+ \Theta(-x^+) + \frac{2x^+}{2+\alpha \kappa \, x^+} \Theta(x^+)~, \quad\quad \tilde{x}^- = x^- - {\alpha}{\kappa} \, \Theta(x^+)/2~,
\end{equation}
applied to the region where $\phi(x^+,x^-)<0$ and where we have rescaled $\alpha\to\alpha/|\phi_h|$. This maps the metric and dilaton in the region $\phi(x^+,x^-)<0$ to the form:
\begin{equation}\label{shockmetric}
ds^2 = \frac{4 d\tilde{x}^+ d\tilde{x}^-}{(1+\tilde{x}^+ \tilde{x}^-)^2} + {2 \alpha}{\kappa} \delta(\tilde{x}^+) (d\tilde{x}^+)^2~, \quad\quad \phi(\tilde{x}^+,\tilde{x}^-) = |\phi_h| \left( \frac{\tilde{x}^+\tilde{x}^--1}{\tilde{x}^+\tilde{x}^-+1}\right)~.
\end{equation}
In this coordinate system, the dilaton exhibits no jump in the dS$_2$ region, whereas the metric has a characteristic $\delta$-function singularity along the shockwave. An analogous procedure can be done for the remaining AdS$_2$ region. 

\begin{figure}[h!]
        \centering
        \subfigure[$\kappa \, \alpha<0$]{
                \includegraphics[scale=0.17]{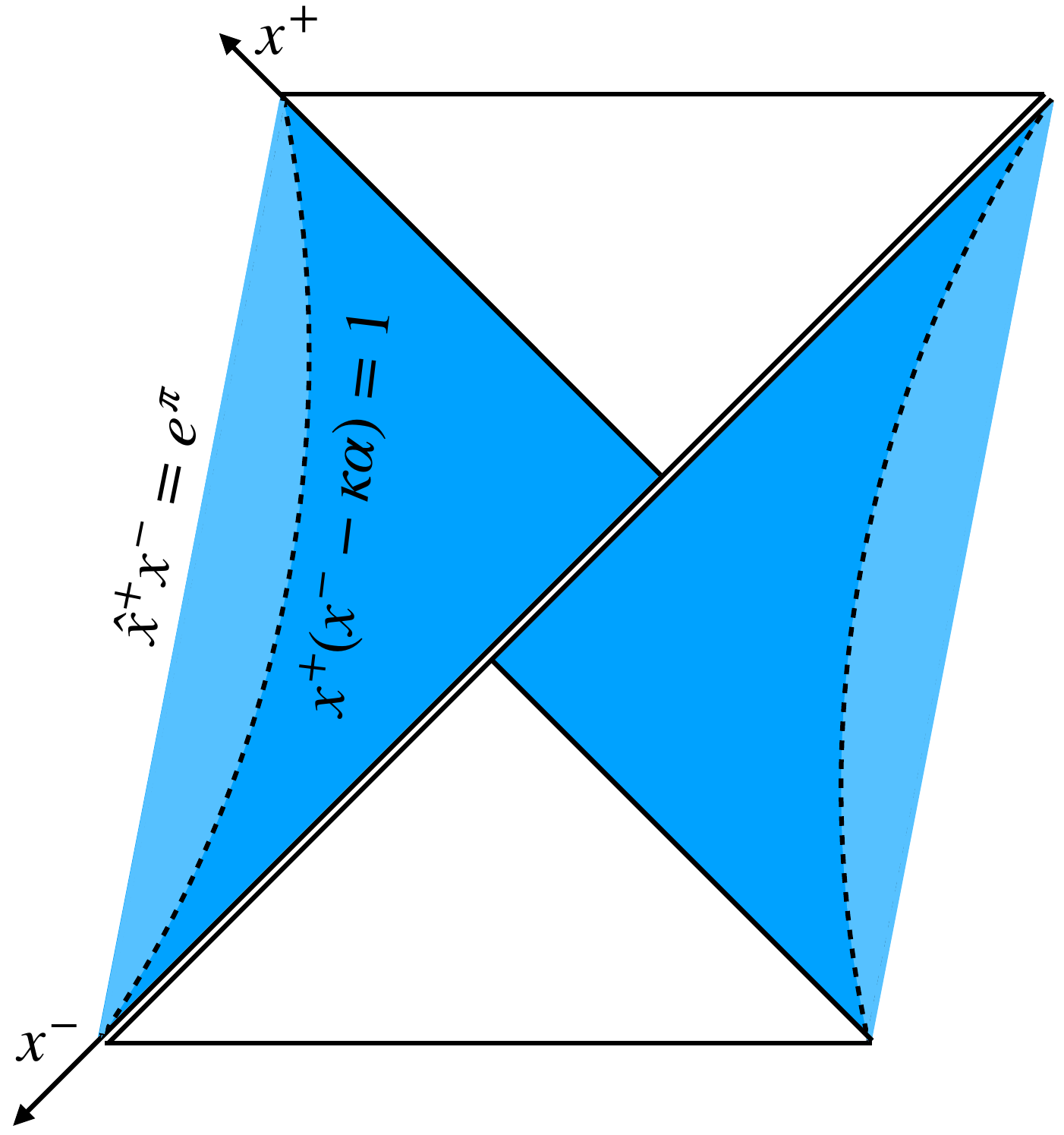}}
                \hspace{5mm}
        \subfigure[$\kappa \, \alpha=0$]{
                \includegraphics[scale=0.17]{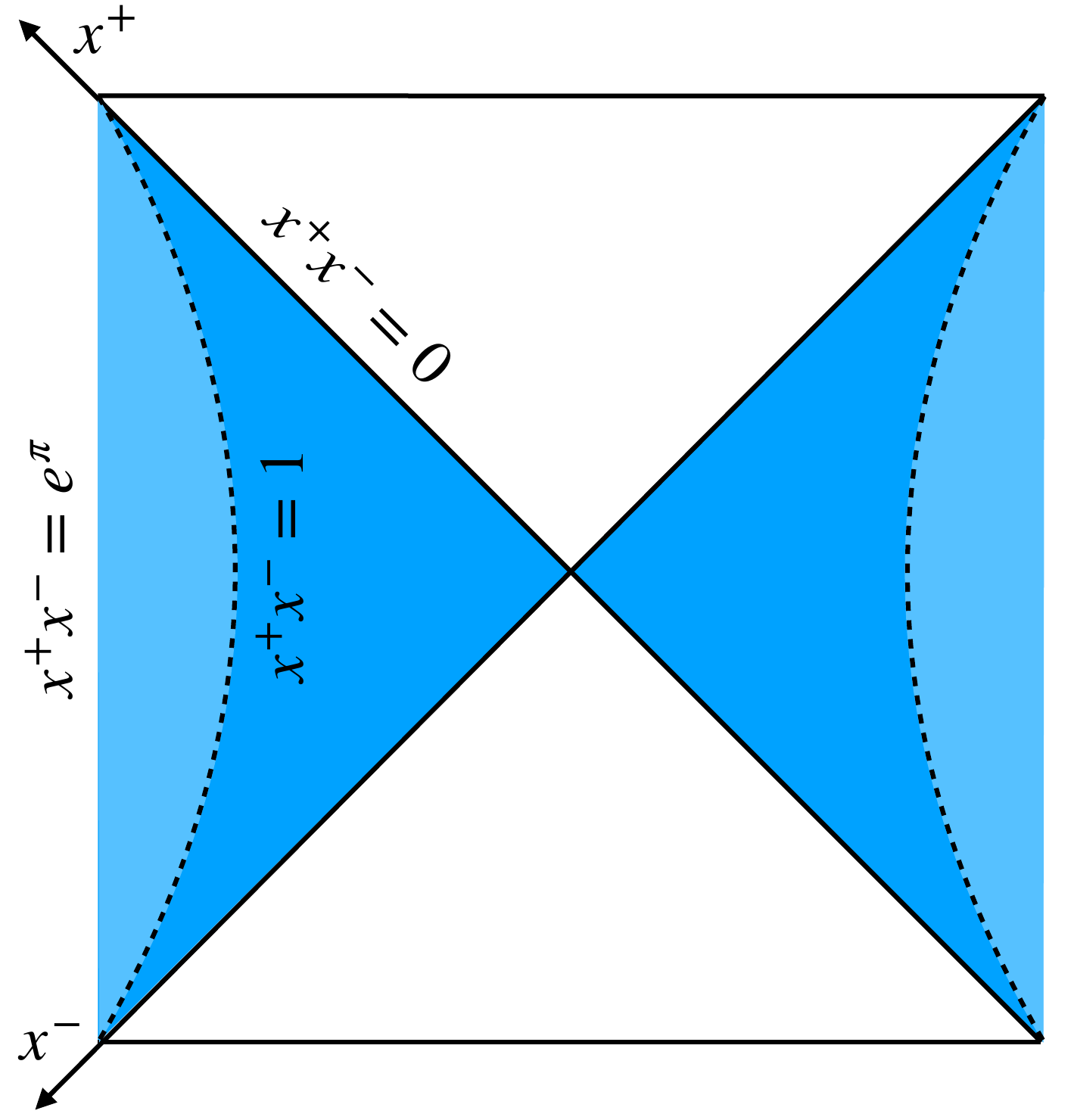}}
                \hspace{5mm}
         \subfigure[$\kappa \, \alpha>0$]{
                \includegraphics[scale=0.17]{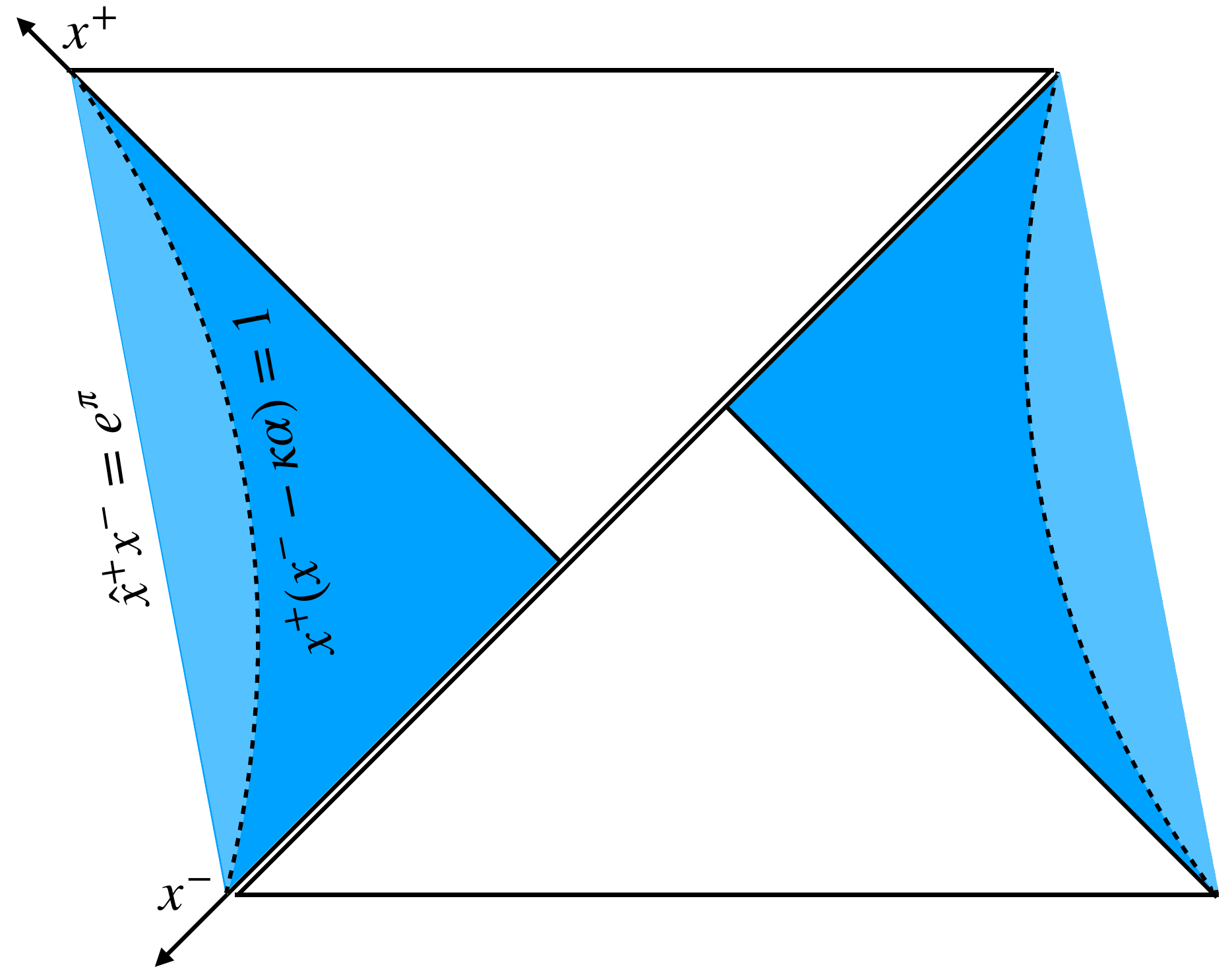}}
                 \caption{{\footnotesize Penrose diagrams for interpolating geometries after a shockwave perturbation.}}
\label{shocks}
\end{figure}

If the integral (\ref{shockI}) is such that $\bar{\phi}+\delta\phi$ becomes zero for some $x^+$, then we must make sure to switch from the dS$_2$ to the AdS$_2$ background geometry. This  defines a curve in the region $x^+>0$:
\begin{equation}\label{curvekruskal}
x^+ \left({x^- - \alpha\kappa}\right)=1~.
\end{equation}
As expected, when $\alpha=0$ this becomes $x^+ x^- = 1$. We schematically depict the geometries in figure \ref{shocks}.

The induced metric $h_{++}$ along the curve is given by:
\begin{equation}
ds^2 = - \frac{4}{(2+\alpha \kappa \, x^+)^2} \left(\frac{dx^+}{x^+}\right)^2~.
\end{equation}
It is also useful to compute the extrinsic curvature along the curve (\ref{curvekruskal}). If we parameterize the curve as $(x^+(u),x^-(u))$, the tangent vector is given by:
\begin{equation}
T^\mu(u) = (\partial_u x^+(u),\partial_u x^-(u))~.
\end{equation}
The normal vector $n_\mu$ obeys $T^\mu n_\mu = 0$ and $n_\mu n^\mu = 1$. Explicitly:
\begin{equation}
n_\mu(u) = \frac{1}{\left| 1+x^+(u)x^-(u)\right|}\left(\sqrt{-\frac{\partial_u x^-(u)}{\partial_u x^+(u)}}, \sqrt{-\frac{\partial_u x^+(u)}{\partial_u x^-(u)}}  \right)~.
\end{equation}
The extrinsic curvature is given by:
\begin{equation}
K_{uu} = T^\mu T^\nu \, \nabla_\mu n_\nu~.
\end{equation}
Evaluating it on the curve (\ref{curvekruskal}), we find that $K_{uu} = 0$. The normal derivative of the scalar along (\ref{curvekruskal}):
\begin{equation}
n^\mu \partial_\mu \phi = 1~.
\end{equation}
To continue the solution to the AdS$_2$ region, we must smoothly glue a constant negative curvature metric across the curve (\ref{curvekruskal}). 

It is convenient to consider a chiral coordinate transformation:
\begin{equation}
\log \hat{x}^+ = \int_{1}^{{x}^+}  \frac{2}{(2+\alpha\kappa u )} \frac{du}{u} = \log \frac{(2+\alpha \kappa){x^+}}{2+\alpha \kappa  {x^+}}~, 
\end{equation}
such that the induced metric on the curve takes the simpler form $ds^2 = -\left(d\hat{x}^+/\hat{x}^+\right)^2$. The range of $\hat{x}^+$ is the positive real axis. Thus, we can smoothly glue (up to first derivatives) half of global AdS$_2$ with coordinates: 
\begin{equation}
e^{2\omega(\hat{x}^+,x^-)} = \frac{1}{\hat{x}^+ x^-}\sec^2 \frac{\log \hat{x}^+ x^-}{2}~,
\end{equation}
along the curve $\hat{x}^+ x^- = 1$. The normal direction then becomes the standard radial direction in global AdS$_2$. The dilaton is given by (\ref{nullAdS2}) with $x^+$ replaced by $\hat{x}^+$.

\subsection{Shapiro time delays and advances}

Notice that for matter obeying the null energy condition, i.e. $\alpha>0$, the shift in the dilaton value due to the stress-tensor in (\ref{phiT}) depends on the sign of the $\kappa$. For $\kappa>0$ ($\kappa<0$), we see that $|\phi|$ increases (decreases) in the region $x^+>0$ due to the shockwave. When $|\phi|$ increases across the shockwave, particles going across the shockwave will experience a Shapiro time-delay. This situation is similar to the case of black holes \cite{Dray:1984ha} -- a flux of positive null energy into the horizon causes a Shapiro time delay. 

What is novel is the case $\kappa<0$ for which $|\phi|$ {decreases} across the shockwave, causing a Shapiro time {advance}. For ordinary black holes, this can only happen if we send in matter that {violates} the null energy condition \cite{Gao:2016bin}. But for a de Sitter type horizon this behaviour is allowed without any violation of the null energy condition. In appendix \ref{dsshock} we connect the $\kappa<0$ behaviour results to a shockwave solution in dS$_3$, for which the dilaton is related to the size of the celestial circle. A more general statement is due to a theorem of Gao and Wald \cite{Gao:2000ga} which states that perturbations obeying the null energy condition make the de Sitter Penrose diagram more vertical, hence allowing access to otherwise causally disconnected regions of space. What is often called the horizon re-entry of super-horizon modes in inflationary cosmology is closely related to these phenomena.

\subsection*{\it{Releasing energy from the AdS$_2$ boundary}}

We would like to make a final remark about the clock appearing in the Lorentzian interpolating geometry: 
\begin{equation}
ds^2 = \begin{cases}
\cos^{-2}\rho \left( - dt^2 + d\rho^2  \right)~, \quad\quad \rho \in (\epsilon,\pi/2)~, \\
 \cosh^{-2}\rho\left( - dt^2 + d\rho^2  \right)~, \quad\quad \rho \in  (-\infty,-\epsilon)~.
\end{cases}
\end{equation}
Imagine that we release a small (in Planck units) energy $\epsilon$ at some early time $t_i \ll - 1$ near the AdS$_2$ boundary of the Lorentzian interpolating geometry. Near the boundary, the timelike isometry is approximately that of the global AdS$_2$ clock. It is also the one that generates time-translations of the inertial clock at $\rho=0$. Thus, the energy in a local frame at $\rho = 0$ is also of order $\epsilon$. The energy $\epsilon$ will consequently enter the dS$_2$ static patch region at the origin, and eventually, near the dS$_2$ horizon at $\rho = -\infty$, the particle will experience a Rindler geometry where the local frame energy will be enhanced. We are interested in how the energy $\epsilon$ depends on $t_i$, when measured in the local frame. 

It is useful to recall the transformation between the static dS$_2$ coordinates $(t,\rho)$ and the global\footnote{As a reminder the global dS$_2$ metric is $ds^2 = -dT^2 + \cosh^2T \, d\varphi^2$ with $\varphi\sim\varphi+2\pi$.} dS$_2$ coordinates $(T,\varphi)$:
\begin{equation}
\cosh^2 T = {1 +\text{sech}^2\rho \, \sinh^2 t}~, \quad\quad \sin^2 \varphi = \frac{\tanh^2 \rho}{1+\text{sech}^2\rho \, \sinh ^2 t}~,
\end{equation}
where we have set $\ell =1$. Translations in $t$ comprise a dS$_2$ isometry. Near the dS$_2$ horizon, $\text{sech} \rho \approx \delta$ with $\delta^2 \ll 1$ and $t \gg 1$, the global and static patch clocks are related in the usual Rindler sense $T \approx \delta \sinh{t}~$. Unlike the relation between the Rindler and Minkowski clock, however, the exponential relation between the global and static dS$_2$ clocks is true only in the regime $\delta e^t \ll 1$. In the limit $\delta e^t \gg 1$, the clocks are linearly related and hence there is no exponential effect. This is in stark contrast to what happens when releasing some energy in a standard AdS black hole with temperature $\beta$. There, the Rindler effect persists for $\beta \ll t \ll \beta \log S_{BH}$ and the local frame energy acquires a factor of $\sim e^{2\pi t / \beta}$ near the horizon such that the connection to the shockwave and the saturation of the Lyapunov exponent follows immediately.

\section{Discussion} \label{discussion}

We end with some general remarks on our results and how they may tie into a holographic picture. Since our solutions have an AdS$_2$ like boundary, the natural candidates for holographic duals will be $(0+1)$-dimensional, i.e. quantum mechanical theories. 

\subsection*{\it dS horizon and chaos}

The chaotic nature of AdS black holes has been heavily explored in recent literature. On the other hand, dS horizons have been less explored from a more modern, holographic perspective (see however \cite{Anninos:2011af,Banks:2012ic,Parikh:2004wh,Goheer:2002vf,Verlinde:2016toy,Dong:2018cuv,Dong:2010pm,Anninos:2010gh,Anninos:2012qw,Anninos:2011zn}). In constructing interpolating solutions between AdS$_2$ and dS$_2$ we have a potential framework to discuss these questions using the standard tools of AdS/CFT, such as correlators at the AdS$_2$ boundary. Using the Hartle-Hawking construction we can build a global state from the Euclidean saddle. To do so, we cut the disk in half such that a state is prepared along the $t=0$ spatial slice. In the absence of any perturbations, the Lorentzian geometry becomes a two-sided interpolating geometry which is smooth across the dS$_2$ horizon. Assuming time-reversal invariance, this solution can be continued all the way to the past. For standard black holes, this state is the thermofield-double state which is built purely out of entangled, non-interacting CFTs. Perturbing this state for black holes, while obeying the null-energy condition, leads to the two sides becoming less connected, and can be viewed as a geometrization of decorrleation due to chaos \cite{Shenker:2013pqa}.

The negative $\kappa$ shockwave solution (\ref{shockmetric}), which behaves qualitatively similar to shockwaves in pure dS, indicates that perturbing the two-sided interpolating geometry leads to more correlation rather than less \cite{Anninos:2017hhn,minic}. Signals may now arrive from the previously causally inaccessible region, something that can only be achieved in the standard black hole case by turning on an interaction between the two CFTs \cite{Gao:2016bin}. From the perspective of a single sided interpolating geometry, perhaps the mechanism is analogous to that of \cite{deBoer:2018ibj}, which might suggest an interesting role of state-dependent operators in the context of dS. 
Perhaps an analogy can be drawn to glassy states which are statically indistinguishable from liquid states but dynamically very different. We view these phenomena as features rather than bugs. They are part of the definition of static patch geometry of dS.


In section \ref{sec_otoc}, we computed the out-of-time ordered correlator for these interpolating geometries, in complete analogy to computations in the AdS$_2$ case \cite{Maldacena:2016upp,Jensen:2016pah, Engel}. The result is surprising, showing an oscillatory correlator (rather than exponential with maximal Lyapunov exponent).  It is worth noting, however, that the propagator $G(u)$ in equation (5.6) contains a piece that resembles the Schwarzian propagator that appears at the boundary of the AdS$_2$ black hole. This might suggest that the maximally chaotic behaviour of the Rindler horizon may be encoded in more subtle correlators. It may also be of interest to study this question for the more general background presented in Appendix \ref{gammaSch}. In that case, it seems that $\gamma \in [-1,1]$ acts as a parameter tuning the behaviour of the out-of-time ordered correlator: for $\gamma>0$, the result is oscillatory; for $\gamma<0$, the correlator is exponential with maximal chaos at $\gamma=-1$; and, at $\gamma=0$, there is an intermediate behaviour where it behaves as a power law, \ie $F(t) \sim t^2$. These features may serve as a guiding principle for the construction of dual microscopic SYK-type models.

\subsection*{\it Role of null-energy violation}

More generally, given the current reassessment of the role of null-energy in the bulk, it may be interesting to reconsider previous attempts to construct dS in a higher dimensional AdS \cite{Freivogel:2005qh,Lowe:2010np}. There, the main stumbling block was that the Raychaudhuri equation plus null-energy obeying matter forbade any dS region to reside in a causally accessible part of the geometry. In view of recent developments \cite{Gao:2016bin,Maldacena:2018lmt}, perhaps one can construct an accessible dS region by turning on weak interactions between the two sides. 


\subsection*{\it dS fragmentation}\label{disc_frag}

In our analysis of anisotropic perturbations we observed the tendency of the region separating positive and negative curvature to separate. In other words, there was only a restricted regime in the space of sources for which one observed a piece of dS in the interior. 

More generally, we might imagine that the interior Euclidean dS region could fragment into several disconnected regions. Perhaps we should interpret these observations as indications that the static patch cannot be arbitrarily sharply defined in and of itself. 

\subsection*{\it Holography of a feature}

Though our discussion was specific to the interpolating geometries of \cite{Anninos:2017hhn}, there is no reason why it could not be applied to more general cases. Understanding the interplay of features in the interior of an asymptotically AdS$_2$ spacetime and the boundary soft modes seems like an interesting general question, and may pertain to broader issues like bulk locality in AdS/CFT.

\subsection*{\it Relation to dS/CFT?}

Finally, it is natural to ask how our approach might be related to the standard dS/CFT picture \cite{Strominger:2001pn,Maldacena:2002vr,Witten:2001kn,Anninos:2011ui}. There, the dual theory resides at the future boundary. From the perspective considered in this paper, the future boundary resides within the horizon of the interpolating geometry and thus, one would have to reconstruct it from the boundary AdS$_2$ degrees of freedom. It would seem that there are far fewer degrees of freedom at the AdS$_2$ boundary to account for those at the future boundary. Perhaps this is an indication that there are unexpected relations among degrees of freedom at the future boundary, as was observed in higher spin models \cite{Anninos:2017eib}.

\section*{Acknowledgements}

We gratefully acknowledge discussions with Frederik Denef, Sean Hartnoll, Ben Freivogel, Juan Maldacena, Rob Myers, Daniel Roberts, Douglas Stanford, Kyriakos Papadodimas, Evita Verheijden, Herman Verlinde, and Erik Verlinde. D.A.'s research is partially funded by the Royal Society under the grant ``The Atoms of a deSitter Universe" and the $\Delta$ ITP. The work of D.A.G. and D.M.H. is part of the $\Delta$ ITP consortium, a program of the NWO that is funded by the Dutch Ministry of Education, Culture and Science (OCW). D.A. would like to thank ICTS, Bengaluru for their kind hospitality during completion of this work. D.A.G. would like to thank the Galileo Galilei Institute for Theoretical Physics for hospitality and ACRI and INFN for partial support during the completion of this work. This project has received funding from the European Research Council (ERC) under the European Union’s Horizon 2020 research and innovation programme (grant agreement No 715656)”

\appendix

\section{Jackiw-Teitelboim gravity with $\kappa<0$}\label{wrongsign}

In this appendix we discuss some solutions to Jackiw-Teitelboim gravity with action:
\begin{equation}
S = S_{top} - \frac{1}{2\kappa} \int d^2x \sqrt{g} \phi \left( R  + 2 \right) - \frac{1}{\kappa} \int_{\partial \mathcal{M}} \sqrt{h} \, \phi \, K~,
\end{equation}
where $S_{top}$ is given by (\ref{stop}). The metric is fixed to be AdS$_2$ due to the dilaton equations of motion. The smooth solution with single boundary is:
\begin{equation}
ds^2 = \frac{d\rho^2 + d\tau^2}{\sinh^2\rho}~, \quad\quad \tau \sim \tau + 2\pi~, \quad\quad \rho\ge0~.
\end{equation}
with $\tau$-independent dilaton:
\begin{equation}
\phi(\rho) = \phi_h \coth \rho~, \quad \phi_h \in \mathbb{R}~.
\end{equation}
There is also a static solution on the Euclidean cylinder:
\begin{equation}\label{globalJT}
ds^2 = \frac{d\rho^2 + d\tau^2}{\cos^2\rho}~, \quad\quad \phi(\rho) = \phi_h \, \tan\rho~,
\end{equation}
where now $\rho \in (-\pi,\pi)$. One can periodically identify $\tau$ with any periodicity in (\ref{globalJT}) without spoiling the smoothness of the solution. 

Since the sign of $\phi_h$ is not fixed, the theory admits solutions with an increasingly negative dilaton near the boundary. As discussed in the main text, at the semi-classical level what we require is that $(\phi_0 + \phi/\kappa) > 0$. The negative $\phi$ solution is equivalent to a positive $\phi$ solution with $\kappa \to -\kappa$. If we view the theory as a dimensional reduction of Einstein-Maxwell theory, the solutions correspond to near-extremal Reissner-Nordstrom black holes. The solutions with negative dilaton correspond to a static, non-asymptotically flat solution with a timelike singularity at the origin which is surrounded by a horizon. Though singular, the `wrong sign' solutions are supported by standard null-energy preserving matter. The boundary dynamics will also be the Schwarzian, at least in the regime $\phi_0^2 \gg (\phi_b/\kappa)^2 \gg 1$. However, the Schwarzian will have the opposite sign corresponding to a `negative' specific heat. From the higher dimensional perspective, this sign indicates that the size of the celestial sphere shrinks toward the UV part of the geometry where time flows the fastest. At a qualitative level, this is also what happens for the static patch of de Sitter space. 

\section{Shockwaves in dS$_3$}\label{dsshock}

In this appendix we discuss a shockwave solution in dS$_3$ and connect it to the analysis in section \ref{shocksec}. 
The following Lorentzian shockwave geometry:
\begin{equation}\label{shockds3}
\frac{ds^2}{\ell^2} =  \frac{4dx^+ dx^-}{\left(1+x^+(x^- + \alpha \Theta(x^+)/2 )\right)^2} + \mu^2 \left(\frac{1-x^+( x^-+ \alpha \Theta(x^+)/2 )}{1+x^+(x^-+ \alpha \Theta(x^+)/2 )} \right)^2 d\varphi^2~,
\end{equation}
with $\varphi \in [0,2\pi]$, $(x^+,x^-) \in \mathbb{R}^2$, solves the three-dimensional Einstein's equations with a positive cosmological constant $\Lambda = +1/\ell^2$ in the presence of a stress energy tensor $T_{++} = \alpha\delta(x^+)$  (where we have set $8\pi G_3 = 1$). The null-energy condition enforces $\alpha>0$. The worldlines of the Northern and Southern static patch observers are at $x^+ x^- = 1$. The future and past boundaries are at $x^+ x^- = -1$. The parameter $\mu^2 \in (0,1]$, with the universe overclosing at $\mu^2=0$ and the pure dS$_3$ universe at $\mu=1$. In the absence of a shockwave the Penrose diagram is that of two conical defects sitting at the North and South poles of dS$_3$. These cause the horizon to be smaller than that of pure dS$_3$, with the limiting case being $\mu^2 = 1$. A useful coordinate transformation is:
\begin{equation}
\tilde{x}^- = x^- + \alpha \Theta(x^+)/2~, 
\end{equation}
for which we have the geometry:
\begin{equation}\label{ds3shock}
\frac{ds^2}{\ell^2} = \frac{4 dx^+ d\tilde{x}^-}{\left(1+ x^+\tilde{x}^- \right)^2} + {\mu^2} \left(\frac{1-x^+\tilde{x}^- }{1+x^+\tilde{x}^-}\right)^2 d\varphi^2 - 2\alpha \delta(x^+)(dx^+)^2~,
\end{equation}
The deformed geometry, with $\alpha \neq 0$, can be viewed as a de Sitter universe with a boosted circular shell from the North pole static patch worldline. We can compare the metric (\ref{ds3shock}) to (\ref{shockmetric}). The absolute value of the dilaton is equivalent to the size of the $\varphi$-circle. Notice that it corresponds to the case $\kappa<0$ since the sign of the $\delta$-function piece is negative. This implies that a light particle traveling near the shock will experience a Shapiro time-advance, allowing it to enter the region of dS$_3$ that was out of causal contact in the absence of the shockwave. 

Analogous solutions can be considered in higher dimensions \cite{Sfetsos:1994xa,Hotta:1992qy,Hotta:1992wb}. The simplest is a geometry connecting two dS$_{d+1}$ regions with a shock that lives exactly on the cosmological horizon. One simply replaces the circle with a $d$-sphere for the $\mu=1$ geometry in (\ref{shockds3}).

\section{Some details for out-of-time-ordered correlators}

\label{app_otoc}
Here we present some explicit results for the ordered and out-of-time ordered 4-points functions in the centaur background. Recall that the full connected 4-points function is given by
\begin{equation}
\left< \left(\frac{\mathcal{B}(u_1,u_2)}{\sin^2 \frac{u_1-u_2}{2}} + \tilde{\mathcal{E}}(u_1,u_2) \right)\left(\frac{\mathcal{B}(u_3,u_4)}{\sin^2 \frac{u_3-u_4}{2}} + \tilde{\mathcal{E}}(u_3,u_4) \right) \right> \,.
\end{equation}
There are then, four different contributions to the correlator, that schematically we name $\langle \mathcal{B} \mathcal{B} \rangle$, $\langle \mathcal{B} \tilde{\mathcal{E}} \rangle$, $\langle \tilde{\mathcal{E}} \mathcal{B}  \rangle$ and $\langle \tilde{\mathcal{E}} \tilde{\mathcal{E}}  \rangle$.

The first contribution can be computed exactly as an analytic function of the four points $u_1,u_2,u_3,u_4$. We can consider first the ordered correlator with $u_4<u_3<u_2<u_1$. Taking $\kappa/\tilde{\phi}_b=1$, this gives
\begin{eqnarray}
& & \langle \mathcal{B}(u_1,u_2)\mathcal{B}(u_3,u_4) \rangle_{ord} =  \pi \, \text{csch} \, \pi  \left(\cosh \left(u_{31}+\pi \right)+\cosh \left(u_{32}+\pi \right)+\cosh \left(u_{41}+\pi \right)+\cosh \left(u_{42}+\pi \right)\right) \nonumber \\
& & + \cot \left(u_{12}/2\right) \left(2 u_{12}+\pi  \, \text{csch} \, \pi  \left(\sinh \left(u_{31}+\pi \right)-\sinh \left(u_{32}+\pi \right)+\sinh \left(u_{41}+\pi \right)-\sinh \left(u_{42}+\pi \right)\right) \right) \nonumber \\
& &+\cot \left(u_{34}/2\right)\left(2 u_{34}+\pi  \, \text{csch} \, \pi \left(-\sinh \left(u_{31}+\pi \right)-\sinh \left(u_{32}+\pi \right)+\sinh \left(u_{41}+\pi \right)+\sinh \left(u_{42}+\pi \right)\right) \right) \nonumber \\
& & + \cot \left(u_{12}/2\right) \cot \left(u_{34}/2\right) \nonumber  \\
& & \left( \pi  \, \text{csch}\, \pi \left(-\cosh \left(u_{31}+\pi \right)+\cosh \left(u_{32}+\pi \right)+\cosh \left(u_{41}+\pi \right)-\cosh \left(u_{42}+\pi \right)\right)-u_{12}u_{34} \right) -4 \,,  \nonumber 
\end{eqnarray}
where we introduced the usual notation $u_{ij} \equiv u_i-u_j$. In the main text, we are interested in the out-of-time-ordered correlator (otoc). In this case, $u_4<u_2<u_3<u_1$, and taking again $\kappa/\tilde{\phi}_b=1$, we obtain:
\begin{multline}
\langle \mathcal{B}(u_1,u_2)\mathcal{B}(u_3,u_4) \rangle_{otoc} = \langle \mathcal{B}(u_1,u_2)\mathcal{B}(u_3,u_4) \rangle_{ord} +  2 \pi u_{32} \cot \left(u_{12}/2\right) \cot \left(u_{34}/2\right)  + \\
 \frac{2 i \pi  e^{-u_2-u_3} \left(e^{u_2}-e^{u_3}\right) \left(e^{u_3} \left(i e^{i u_1}+e^{i u_2}\right) \left(i e^{i u_3}+e^{i u_4}\right)+e^{u_2} \left(e^{i u_1}+i e^{i u_2}\right) \left(e^{i u_3}+i e^{i u_4}\right)\right)}{\left(e^{i u_1}-e^{i u_2}\right) \left(e^{i u_3}-e^{i u_4}\right)} \,. 
\end{multline}
Given this formula, it is straightforward to compute the contribution of this term to the real-time 4-points function that appears in equation (\ref{ef_of_t}) in the main text, that results in (\ref{nochaos}). Moreover, a completely analogue calculation using the propagator that corresponds to the AdS$_2$ black-hole case gives the saturation of the Lyapunov exponent for quantum chaotic systems \cite{Maldacena:2016hyu,Maldacena:2016upp}.

\begin{figure}[h!]
\centering
        \subfigure[$\langle \tilde{\mathcal{E}} \mathcal{B}  \rangle$]{
                \includegraphics[scale=0.36]{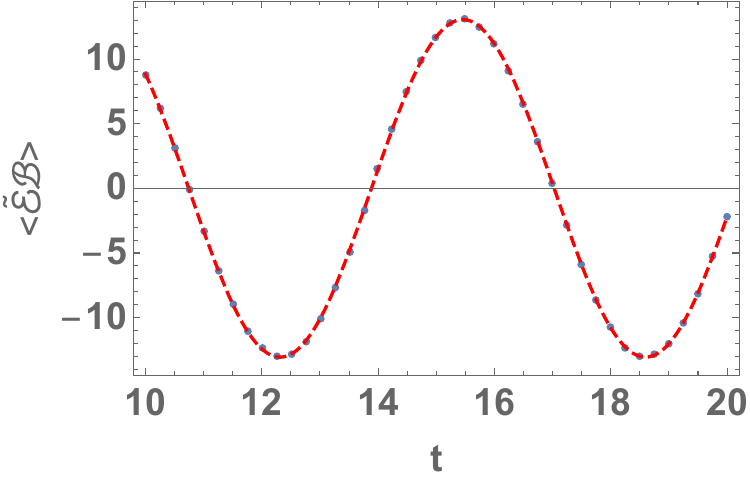}}
                \hspace{5mm}
        \subfigure[$\langle \mathcal{B} \tilde{\mathcal{E}} \rangle$]{
                \includegraphics[scale=0.36]{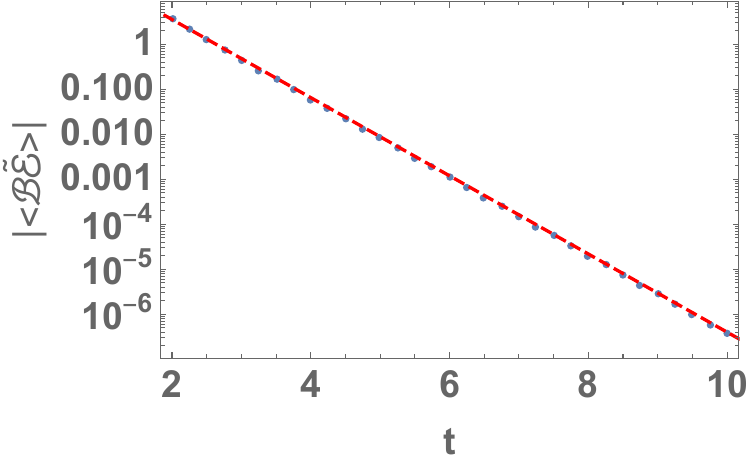}}\quad
                \hspace{5mm}
         \subfigure[$\langle \tilde{\mathcal{E}} \tilde{\mathcal{E}}  \rangle$]{
                \includegraphics[scale=0.36]{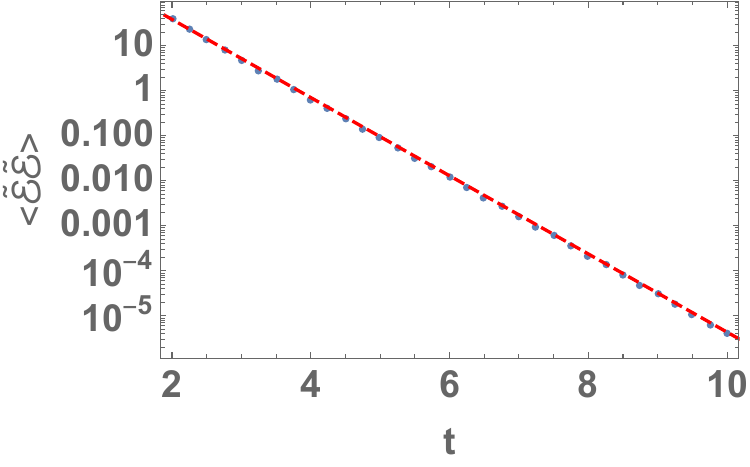}}
\caption{\footnotesize Three contributions to the otoc in the case of the interpolating geometry. The blue dots correspond to the numerical solution while the dashed red lines correspond to the best fit by functions of the form $A e^{-2 B t}$ for the decaying terms and $A \cos (B t+ C)$ for the oscillating one. We set $\tilde{\beta}=2\pi$ and $\kappa/\tilde{\phi}_b =1$. $A,B,C$ are constants to be fitted. In all cases, the best fit gives $B\approx1$.}
\label{numeric}
\end{figure}

In the case of the interpolating geometries considered in the main text, we have three additional contributions to the out-of-time-ordered four-point function. As argued in the main text, these contributions either oscillate, or are exponentially suppressed as a function of real time $t$. Here, we provide numerical evidence to support this claim. 
In figure \ref{numeric} we display the full contribution to $\langle \mathcal{B} \tilde{\mathcal{E}} \rangle$ and $\langle \tilde{\mathcal{E}} \tilde{\mathcal{E}}  \rangle$ -- they behave as $\sim e^{-4\pi t/\tilde{\beta}}$. The $\langle \tilde{\mathcal{E}} \mathcal{B}  \rangle$-terms oscillate with frequency $2\pi/\tilde{\beta}$. None of these contributions grow exponentially with time.
\section{The $\gamma$-theory and $\gamma$-Schwarzian}\label{gammaSch}

In this appendix we generalise the dilaton potential to $V(\phi) = 2(|\phi-\phi_0|-\phi_0)$. The solution for $\phi<\phi_0$ is now given by:
\begin{equation}\label{pos}
ds^2 = \frac{d\rho^2+d\tau^2}{\cosh^2\rho}~, \quad\quad \phi(\rho) =  -\phi_h \tanh\rho~,
\end{equation}
with $\phi_h<0$. Smoothness requires $\tau\sim\tau+2\pi$. At $\tanh \rho_c = -{\phi_0}/{\phi_h}$,
the geometry interpolates to one with negative curvature. 

The general negative curvature solution is given by:
\begin{eqnarray}\label{gammametric}
ds^2 = \frac{\gamma}{\sin^2 \sqrt{\gamma}R} (dR^2+d\tau^2)~, \quad\quad \phi (R)= \alpha  \cot \left(\sqrt{\gamma } R\right)+2\phi_0~.
\end{eqnarray}
For $\gamma>0$ we have $R \in (0,\pi/\sqrt{\gamma})$ whereas for $\gamma<0$ we have $R \in (0,\infty)$. Locally we can rescale $R$ and $\tau$, and thus (\ref{gammametric}) is diffeomorphic to the standard hyperbolic metric on the disk. However, since we are fixing the periodicity of $\tau$ the above geometry is globally distinct from the standard hyperbolic metric. Indeed, for $\gamma<0$ the geometry (\ref{gammametric}) contains a conical defect with deficit angle $\Delta\phi = 2\pi (1-\sqrt{-\gamma})$. Thus, for $\gamma=-1$ there is no deficit and for $\gamma<-1$ there is a conical surplus. As $\gamma \to 0^-$, the geometry tends to 
\begin{eqnarray}\label{gammametric2}
ds^2 = \frac{d\tau^2 + dR^2}{R^2}~, \quad\quad \tau\sim\tau+2\pi~.
\end{eqnarray}
Here the geometry develops an infinite throat as $R \to \infty$. 
\begin{figure}[h!]
        \centering
        \subfigure[$\phi_0<0$]{
                \includegraphics[scale=0.5]{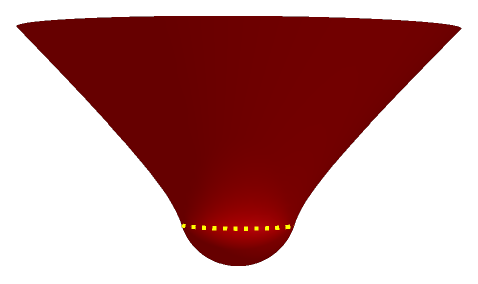} \label{kappaless}}
                \hspace{5mm}
        \subfigure[$\phi_0=0$]{
                \includegraphics[scale=0.5]{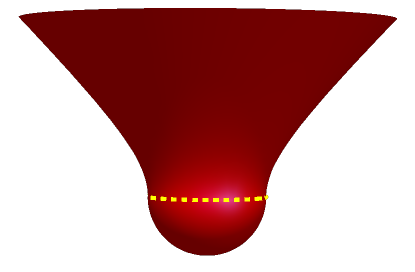}}\quad
                \hspace{5mm}
         \subfigure[$\phi_0>0$]{
                \includegraphics[scale=0.39]{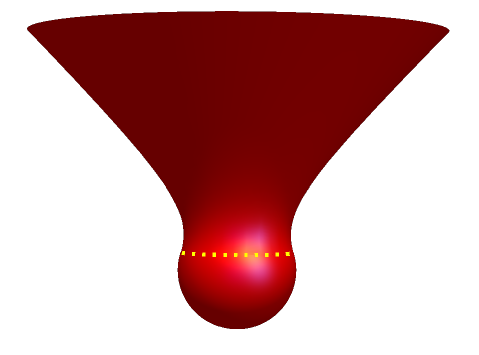} \label{kappamore}}
                 \caption{\footnotesize The different interpolating geometries as a function of $\phi_0$. In general, negative $\phi_0$ solutions contain a smaller part of the sphere than the $\phi_0=0$, while $\phi_0>0$ contain a larger part. }
\label{Trrfigs}
\end{figure}

We need to glue the two geometries (\ref{pos}) and (\ref{gammametric}) so that both the metric and the dilaton are smooth up to first derivatives. This is achieved by gluing the negative curvature solution at a specific $R_g$ given by
\begin{eqnarray}
\tan \sqrt{\gamma}R_g = {\sqrt{\frac{2 \gamma }{1-\gamma }}} \,, \quad \quad \gamma = 1-{2 \phi_0^2}/{\phi_h^2} \,.
\end{eqnarray}
This is valid only for positive $\rho_c$ and positive $\gamma$. This completes the smooth gluing at the interpolating region. Finally, boundary conditions must be imposed near the AdS$_2$ boundary of (\ref{gammametric}) to complete the Dirichlet problem. We give a schematic depiction of the different geometries in figure \ref{Trrfigs}.

\subsubsection*{{\it Case (i): $\phi_0>0$}}

\begin{figure}[h!]
 \begin{center}
\includegraphics[width=2cm]{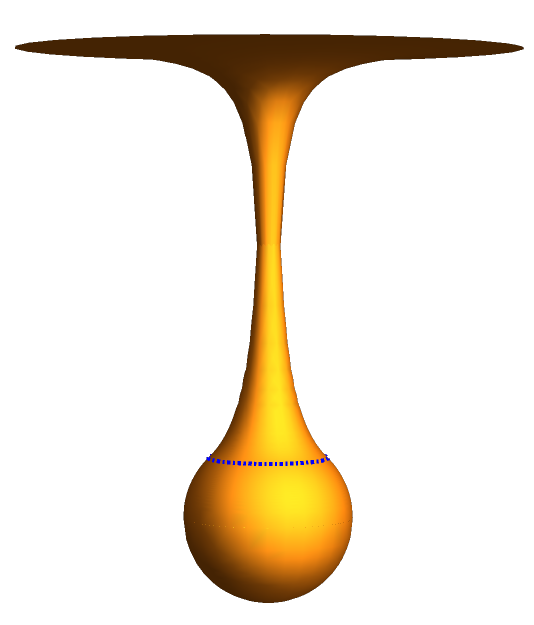}
\caption{{\footnotesize A sketch of the interpolating geometry close to $\gamma \to 0^+$ with $\phi_0>0$. As $\gamma$ approaches to zero, the interpolating geometry develops an infinitely large throat until it gets disconnected for $\gamma>0$.}}\label{disconnected}
\end{center}
\end{figure}

For $\phi_h>0$ one finds a pure AdS$_2$ solution, that for $\kappa>0$ ($\kappa<0$) will be the dominant (sub-dominant) saddle. For $\phi_h<0$ one finds that $\rho_c>0$ and an interpolating geometry exists so long as $\rho_c \lesssim 0.88$. In these cases, the interpolating geometry will have a greater portion of the sphere compared to the $\phi_0=0$ case. As we move along the $\phi_h$ negative values, we change the interpolating radius. The limiting case $\phi_h = -\sqrt{2} \phi_0$ corresponds to a geometry with $\gamma \to 0$, i.e. one that is about to be disconnected. This is shown in figure \ref{disconnected}. For greater values of $\phi_h$, only the AdS$_2$ geometry exists. So, for $\phi_0>0$, only interpolating solutions with $\gamma \in [0,1]$ exist. 

\subsubsection*{{\it Case (ii): $\phi_0<0$}}

For $\phi_0<0$, the interpolating geometry contains a smaller piece of two-sphere. The interpolating radius $\rho_c$ is always negative and it is possible to have $\gamma \in [-1,1]$ in this case. A transition occurs at $\gamma=0$, where $\phi_h = \sqrt{2} \phi_0$, but the interpolating geometry remains smooth and connected. In the limit of $\phi_h \rightarrow \phi_0$, the sphere is lost completely. Note that these solutions do not have a zero temperature limit. 

\subsection{The $\gamma$-Schwarzian theory}

It is interesting to consider the boundary mode actions that emerge for boundary fluctuations in the geometries (\ref{gammametric}). In equation (\ref{bdyaction}), we saw that for global AdS ($\gamma = 1$) the boundary action is a Schwarzian plus a kinetic term that has the opposite sign to the usual AdS$_2$-black hole. This, in turn, generates an oscillating OTOC as opposed to the black hole where the OTOC grows exponentially and saturates the chaos bound.  

It is interesting to see how the Schwarzian action is modified by considering boundary fluctuations of a general negative curvature geometry (\ref{gammametric}). We find:
\begin{equation}\label{bdyaction_aapendix}
S_{bdy} =  \frac{\phi_b}{\kappa} \int du  \left( \frac{\gamma}{2} \, (\partial_u \tau (u))^2 -  \text{Sch}\left[\tau(u),u  \right] \right)~.
\end{equation}
Here $\tau(u)$ maps $u$ to a circle of period $2\pi$. Note that $\gamma=1$ gives the global AdS action already discussed in section \ref{schwartzian}, while $\gamma=-1$ gives the standard Schwarzian action for the hyperbolic disk. The above action has a saddle at $\tau(u) = 2\pi u/\tilde{\beta}$, such that $u \sim u + \tilde{\beta}$. Considering an expansion as $\tau(u)=2\pi u/\tilde{\beta} + \delta\tau(u)$, the action for this $\gamma$-Schwarzian becomes (to quadratic order in $\delta\tau$),
\begin{eqnarray}
S_{bdy} = \frac{\phi_b}{2 \kappa} \int_0^{\tilde{\beta}} du \left( \gamma \, \delta\tau'(u)^2 + \frac{\tilde{\beta}^2}{4\pi^2} \,  \delta\tau''(u)^2 \right) \,.
\end{eqnarray}

\end{document}